\documentclass[useAMS,usenatbib]{mn2e}

\usepackage{graphicx}
\usepackage{epsfig}

\newcommand{\ha}{H$\alpha~$}

\def\vhel{\ifmmode{V_{{\rm HEL}}}\else{$V_{{\rm HEL}}$}\fi}
\def\vsys{\ifmmode{V_{\rm sys}}\else{$V_{\rm sys}$}\fi}
\def\kms{\ifmmode{~{\rm km\,s}^{-1}}\else{~km~s$^{-1}$}\fi}
\def\vlsr{\ifmmode{v_{\rm lsr}}\else{$v_{\rm lsr}$}\fi}

\def\ltsim{\ifmmode\stackrel{<}{_{\sim}}\else$\stackrel{<}{_{\sim}}$\fi}
\def\gtsim{\ifmmode\stackrel{>}{_{\sim}}\else$\stackrel{>}{_{\sim}}$\fi}

\begin{document}

\title[Towards a free-free template for CMB foregrounds] {Towards a free-free
template for CMB foregrounds}

\author[C. Dickinson et al.]
{C. Dickinson,$^{1}$\thanks{E-mail: cdickins@jb.man.ac.uk}
R. D. Davies,$^{1}$ and R. J. Davis$^{1}$ \\
$^{1}$Jodrell Bank Observatory, Dept of Physics \& Astronomy,
University of Manchester, Macclesfield, Cheshire SK11 9DL UK. \\}

\date{Received
**insert**; Accepted 21st January, 2003}

\pagerange{\pageref{firstpage}--\pageref{lastpage}} \pubyear{}

\maketitle
\label{firstpage}

\begin{abstract}

A full-sky template map of the Galactic free-free foreground emission
component is increasingly important for high sensitivity CMB
experiments. We use the recently published \ha data of both the
northern and southern skies as the basis for such a template.

The first step is to correct the \ha maps for dust absorption using
the 100 $\mu$m dust maps of Schlegel, Finkbeiner \& Davis
(1998). We show that for a range of longitudes, the Galactic latitude
distribution of absorption suggests that it is 33 per cent of the full extragalactic absorption. A reliable absorption-corrected \ha map
can be produced for $\sim 95$ per cent of the sky; the area for which
a template cannot be recovered is the Galactic plane area $|b| < 5^{\circ}$, $l=260^{\circ}-0^{\circ}-160^{\circ}$ and some isolated
dense dust clouds at intermediate latitudes.

The second step is to convert the dust-corrected \ha data into a
predicted radio surface brightness. The free-free emission formula is revised to give an accurate expression (1 per cent) for the radio
emission covering the frequency range 100 MHz to 100 GHz and the
electron temperature range 3000 to 20000 K. The main uncertainty
when applying this expression is the variation of electron temperature
across the sky. The emission formula is verified in
several extended H{\sc ii} regions using data in the range 408 to 2326
MHz.

A full-sky free-free template map is presented at 30 GHz; the scaling to
other frequencies is given. The Haslam et al. all-sky 408 MHz map of the sky
can be corrected for this free-free component, which amounts to
a $\approx 6$ per cent correction at intermediate and high latitudes, to provide a pure
synchrotron all-sky template. The implications for CMB experiments are discussed.

\end{abstract}



\section{Introduction}
\label{intro}

Current CMB experiments are sensitive enough to measure the primordial fluctuations
which have an amplitude in the range of $\approx
20-100~ \mu$K over the $\ell$-range $10-2000$ which corresponds to angular scales of $\approx 10^{\circ} -
10$ arcmin (Hanany et al. 2000; Mauskopf et al. 2000; Padin
et al. 2001; Halverson et al. 2002; Scott et al. 2002). The angular
power spectrum of these fluctuations contains a wealth of
cosmological information.  One of the crucial factors in determining
an accurate power
spectrum of these fluctuations is understanding and removing the foreground
contamination. The amplitude of the foreground signal depends on
frequency, angular scale and region of sky being observed. CMB
foregrounds comprise point sources and diffuse Galactic
foregrounds. Point sources are a particular problem at smaller angular
scales ($\la 30$ arcmin) , while diffuse foreground structures become
dominant on larger angular scales. Here we concentrate on the problem of
diffuse Galactic foregrounds at frequencies below $< 100$ GHz, a
frequency range which is used in current and upcoming CMB experiments
such as the {\it MAP} and {\it Planck} satellites.  

\subsection{Galactic foregrounds}
\label{gal_foregrounds}

The diffuse Galactic foreground has 3 (possibly 4)
components. i) {\it Synchrotron} emission from relativistic electrons spiralling in the
Galactic magnetic field dominates at frequencies below 1 GHz
with a spectral index ($T \propto \nu^{- \beta}$) of $\beta \approx
2.7-3.2$ (Davies et al. 1996). It
is traced by low-frequency surveys such as those of Haslam et
al. (1982), Reich \& Reich. (1988) and Jonas et al. (1998) where it is seen to extend well
above the plane - e.g. the North Polar Spur which extends to Galactic
latitude $b
\approx 80^{\circ}$. ii) {\it Free-free} bremsstrahlung emission from thermal
electrons has a flatter spectral index of $\beta \approx 2.1$. It is the dominant foreground at frequencies between $\nu =
10-100$ GHz, where microwave CMB experiments operate (e.g. DASI, CBI, VSA,
{\it MAP}, {\it Planck}). The optical \ha line is a good tracer of free-free
emission although it requires corrections for dust absorption. iii)
The {\it vibrational} emission from thermal dust is
dominant at frequencies above $\sim 100$ GHz. This is thermal emission
from warm dust and is well-traced at $\lambda \sim
100~ \mu$m where it has its peak. iv) An {\it anomalous}
component has recently been discovered (Kogut et al. 1996; Leitch et
al. 1997; de Oliveira-Costa et al. 1997,1998,1999,2000,2002;
Finkbeiner et al. 2002). Draine \& Lazarian
(1998) proposed that this could be due to $spinning$ dust grains
emitting in a 1-2 octave band centred at $\sim 20$ GHz. It was
first believed to be due to free-free emission (Kogut 1996), but this has
been ruled out primarily by the lack of associated \ha emission. 

The separation and quantifying of the individual CMB foreground
components is a continuing challenge. Each component is of interest
in its own right; its angular distribution and spectrum are basic
parameters. They all contribute as foregrounds to the CMB and their
removal is necessary to realize the full potential of current high
sensitivity CMB surveys where $C_{l}$ values are to be measured to an
accuracy of a few percent. Such separations are best made using
templates for each component. With the advent of the \ha survey, such a
template is now available for each of the 4 proposed
components. Although some approaches to separation of components are
being made which reduce the assumptions about the template parameters
(e.g. Baccigalupi et al. 2000; Vielva et al. 2001)
they still face difficulties where the components are correlated. For
example all the foregrounds are quasi-correlated as they increase
strongly towards the Galactic plane. Also, the \ha and (spinning) dust
emissions show considerable correlation over most of the sky (see Section
\ref{discussion}); this was at the root of the original
misunderstanding of the anomalous component. If one has a good
free-free template as we derive here, the other 3 foreground
components can be well-characterized except possibly at the lowest
Galactic latitudes.

\subsection{Large-area \ha surveys}
\label{ha_surveys}

Extensive filamentary \ha emission regions, of galactic origin,
were first found by Meaburn (1965, 1967) extending to high Galactic 
latitudes; Sivan (1974) demonstrated the wealth of \ha emission at
low-intermediate latitudes but at a relatively low sensitivity of $\approx
15~R$; we note that 1 Rayleigh $(R) \equiv 10^{6}/4\pi $
photons s$^{-1}$ cm$^{-2}$ sr$^{-1} \equiv 2.41 \times 10^{7}$ erg
s$^{-1}$ cm$^{-2}$ sr$^{-1} \equiv 2.25~ {\rm cm}^{-6} {\rm pc}$ for
$T_{e}=8000$ K gas. Several large and very sensitive \ha surveys are now well
under way. The most relevant surveys are listed in Table \ref{halpha_surveys.table}.
\begin{table*}
\begin{center}
\caption[Current \ha surveys]{Current \ha surveys relevant to CMB observations}
\begin{tabular}{||l|l|c|c|c|c||}
\hline

Survey &Type &Field of View& Resolution &Sensitivity/pixel &Sky \\ 

 & &(degrees) &(arcmin) &($R$) &coverage  \\ \hline \hline

WHAM &Fabry-Perot &1 &60 &$\approx 0.05$ &$\delta \ge -30^{\circ}$  \\
 &Spectroscopy & & & &  \\ \hline

VTSS &Narrow-band &10 &1.6 &$\approx 1$ &$\delta \ge -15^{\circ}$ \\
 &Imaging & & & &  \\ \hline

SHASSA &Narrow-band &13.6 &0.8 &$\approx 2$ &$\delta \le
+15^{\circ}$  \\ 

 &Imaging & & & &  \\ \hline

MWFC &Narrow-band &32 &7 &$\approx 1$ &Selected  \\

 &Imaging & & & &fields  \\ \hline

AAO/ &Narrow-band&5.5 &0.02 &$\approx 5-10$ &$|b| \le 10^{\circ}$  \\

Schmidt &Imaging & & & &  \\ \hline
\label{halpha_surveys.table}
\end{tabular}
\end{center}
\end{table*}
In the northern sky covering declinations $\delta \ge -30^{\circ}$ the
Wisconsin H Alpha Mapper project (WHAM) is the
most sensitive \ha survey to-date, but is restricted to an angular resolution of
$\approx 1^{\circ}$ (Reynolds et al. 1998; Haffner 1999). The final sensitivity
in each field is $\approx 0.05~R$ and the data are calibrated to an
accuracy conservatively estimated at 10 per cent. WHAM is a dual
etalon Fabry-Perot spectrometer with a velocity
resolution of 12 km s$^{-1}$. The spectra can be used to remove the
geocoronal \ha emission from the Earth's upper atmosphere which varies
in strength from 2-13 $R$; it is much brighter than
the Galactic signals at high Galactic latitudes.  

The ongoing Virginia Tech Spectral-line Survey (VTSS) is a complimentary
narrow-band imaging survey of the northern sky covering $\delta \ge -15^{\circ}$ (Dennison, Simonetti \&
Topsana 1998). It has a resolution of 1.6 arcmin in each $10^{\circ}$ diameter
field, which will be good enough for almost all CMB experiments as
most of the cosmological information is contained on scales between
$10^{\circ} - 10$ arcmin. However, on larger angular scales, it will most likely be limited by the
geocoronal emission which appears as a time-varying background signal
of unknown level. Star residuals which have not been subtracted
correctly can also be a problem. In combination with the WHAM survey, this survey will be a powerful tool.

The recently published (Gaustad et al. 2001) Southern H Alpha Sky
Survey Atlas (SHASSA) covers the
southern sky $\delta \le +15^{\circ}$ at an angular resolution of 0.8
arcmin in each $13^{\circ}$ field. The sensitivity reaches $2~R$ limited by geocoronal
emission and star residuals which have been partially removed by median
filtering. Smoothing this survey to a resolution of 4 arcmin allows features of about
$0.5~R$ to be detected. The intensity calibration is accurate to about
9 per cent. 

Two other \ha surveys are in progress. The Manchester Wide-Field
Camera (MWFC) can observe \ha with a $32^{\circ}$ field
of view with 7 arcmin resolution (e.g. Boumis et al. 2001) and a
sensitivity of $\approx 1~R$. This instrument is particularly
sensitive to large scale features ($\ga 1^{\circ}$) which may be
missed by cameras with smaller fields. We have recently observed several
selected fields to look for diffuse \ha emission, both in the northern
and southern hemispheres. The AAO/UK Schmidt \ha survey has high angular
resolution (1 arcsec) but is restricted to the Galactic plane
at relatively low sensitivity (Parker \& Phillips 1998) as given in
Table \ref{halpha_surveys.table}.  

\subsection{This paper}

In this paper we describe the steps taken to derive a free-free
emission template from recently published \ha surveys. In Section
\ref{dust_absorption} we determine the absorption of the Galactic \ha
emission by dust based on the 100 $\mu$m dust template given by
Schlegel, Finkbeiner \& Davis (1998), hereafter SFD98; a statistical estimate is made of the relative distribution of the
\ha emission and the dust in the line of sight. Section
\ref{halpha_conversion} summarizes the expressions required to convert
the dust-corrected \ha emission into microwave emission at frequencies
relevant to current CMB experiments. The uncertainties in the relation
are evaluated. This relationship is tested in Section \ref{obs_tests}
using available radio surveys. The final free-free template is
presented in Section \ref{free-free_template}. The 408 MHz all-sky map of Haslam
et al. (1982) corrected for free-free emission is derived in Section
\ref{full_synch}; it is the best available synchrotron template. We
discuss the implications of the new template for upcoming CMB
experiments and for interstellar medium (ISM) studies in Section
\ref{discussion}. The final conclusions of this work are given in
Section \ref{conclusions}.


\section{Correction of \ha for dust absorption}
\label{dust_absorption}

\subsection{The Galactic distribution of absorption}

It is clear that the use of \ha surveys as a template for
free-free emission will be jeopardized in the Galactic plane where
visual absorption is typically 1 magnitude per kiloparsec in the local
arms and where the total absorption to the Galactic centre is $\approx
20$ magnitudes. At intermediate latitudes the absorption broadly
follows a cosecant law with a vertical slab half-thickness of 0.1 to
0.2 blue magnitudes. Such a cosecant law was widely used in
extragalactic astronomy. In order to take account of the known
structure in the obscuration, the line integral of H{\sc i} combined
with a factor derived from the Shane-Wirtanen galaxy counts was
introduced by Burstein \& Heiles (1978). This procedure took account
of the possibility that the gas-to-dust ratio might not be constant
and the fact that not all hydrogen is in the neutral atomic form. The
Burstein-Heiles approach is limited by the angular resolution of H{\sc
i} all-sky surveys which is currently $\sim 30$ arcmin.

A new approach which provides a resolution of 6.1 arcmin and
has a complete all-sky coverage is offered by SFD98 who use far
infra-red data from the {\it COBE-DIRBE} and {\it IRAS} satellites. This combination of data at a range of
FIR wavelengths enabled a good zero level to be established, an
adequate removal of zodiacal light and a more effective destriping
than previously. Also, by comparing the {\it DIRBE} 240 $\mu$m and {\it IRAS} 100
$\mu$m to derive a dust temperature, an estimate could be made of the
dust column density, $D^{T}$, measured in units
of MJy sr$^{-1}$, in terms of a 100 $\mu$m
surface brightness at a fixed temperature of 18.2 K. This correction to a fixed temperature can only be made on the
angular scale of the {\it DIRBE} observations, namely $0^{\circ}\!.7$, and
consequently is not strictly true on the $6.1$ arcmin scale of the
$D^{T}$ dust maps. The dust temperatures are typically in the range
$17-21$ K which corresponds to a correction of up to a factor of $\sim
5$ in the dust column density (see SFD98). 

\subsection{Derivation of a \ha absorption template}
\label{A_halpha}

We will adopt the $D^{T}$ dust template as the indicator of dust
absorption. This then requires a conversion factor to estimate the absorption at 656.3 nm, the \ha wavelength. SFD98 use the
$(B-V)$ colours of some 470 galaxies at a wide range of Galactic
coordinates to derive a best fit correlation with the $D^{T}$ value
at the position of each galaxy. They find the $E(B-V)$ colour can be
estimated from $D^{T}$ with the expression $E(B-V) = (0.0184 \pm
0.0014) D^{T}$ magnitudes and claim that the reddening estimated in
this way has a standard deviation of 16 per cent at any given position.

In order to estimate the absorption at the \ha wavelength we use the parametric
extinction curve for optical wavelengths given by O'Donnell (1994) and
find the Galactic absorption at \ha to be $A ({\rm H}\alpha) = 0.81
A(V)$. Assuming the dust is characterized by a reddening value $R_{V} = A(V) / E(B-V) =
3.1$, this leads to an absorption at \ha of 
\begin{equation}
\label{dust_absorption.eqn}
A({\rm H}\alpha) = 2.51 E(B-V) = (0.0462 \pm 0.0035) D^{T} {\rm mag}
\end{equation}
\noindent where $D^{T}$ is in units MJy sr$^{-1}$. The values given in table 6
of SFD98 are 6 per cent higher than this,
corresponding to a higher value of $R_{V}$.

A range of reddening laws, characterized by different values
of $R_{V}$, are found in different directions in the Galaxy. The
different laws are thought to derive from different grain chemistry
and size distributions. Values of $R_{V}$ range from 2.5 to 5, with
the higher values often in denser dust clouds. A value of 3.1 is
widely adopted as representative of the diffuse ISM in directions away
from dense dust clouds. However, it should be emphasized that this is
only an average value and that any line of sight will have a scatter
about this value. Putting all these statistical factors together we
consider that the relationship between $A({\rm H}\alpha)$ and $D^{T}$
is accurate to about 20 per cent.

\subsection{The absorption of Galactic \ha}
\label{abs_gal}

The expression given in the preceding section applies to absorption of
extragalactic objects at optical wavelengths. Here we are considering
Galactic \ha emission which is mixed with the dust. At first
sight it might be argued that since the dust and gas are uniformly
mixed, then the ionized gas and the dust would also be uniformly
mixed. The correlation between gas and dust will be discussed further
in Section \ref{discussion}. The major
part of the \ha seen from our position in the Galaxy  is the
result of ionization by the UV radiation field from young stars
embedded in the
dust plus the contribution  from nearby regions of localized recent
star formation. Such regions include the Gould Belt system which
reaches to Galactic latitudes of $40^{\circ}$ or more.

We seek a first-order absorption formula for correction of the
\ha emission based on the dust distribution given by SFD98. If
the ionized gas were uniformly mixed with the dust then the average
absorption would be half the extragalactic value given in Section
\ref{A_halpha} on the assumption that the dust absorption is optically
thin (say $\le 0.5$ mag). We define $f_{d}$,
the effective dust fraction in the line of sight actually absorbing
the \ha$\!\!$ so that the actual absorption of \ha is given by $f_{d}
\times A({\rm H}\alpha)$, where $A({\rm H}\alpha)$ is the full
extragalactic absorption. For example, uniformly mixing of ionized gas
and dust corresponds to $f_{d} = 0.5$. There would be a
systematic modification of this value if the z-distribution of
the dust and the ionized gas were not the same. An example would be the
ionization produced by the interstellar radiation field propagating
away from the plane and ionizing the ``under''-side of the gas clouds
which were themselves optically thick to the UV ionizing
radiation. This would lead to a narrower z-distribution of \ha
emission than dust, which is assumed to be well-mixed with the gas, and
to a reduced value of $f_{d}$. 

We are now in the position to test the uniform mixing hypothesis by comparing at
low and intermediate latitudes the z-distribution of neutral gas (H{\sc i}),
ionized gas (H$\alpha$) and dust.

\subsection{Galactic latitude scans}  
\label{latitude_scans}

By plotting the \ha dependence on Galactic latitude $b$, and comparing this with the
unabsorbed gas indicators of H{\sc i} and FIR dust ($D^{T}$), we can estimate
the \ha absorption as a function of $b$ and determine a best-fitting
value of $f_{d}$, the effective dust fraction producing absorption.

\begin{figure*}
\setlength{\unitlength}{1mm}
\begin{picture}(80,120)
\put(0,0){\includegraphics{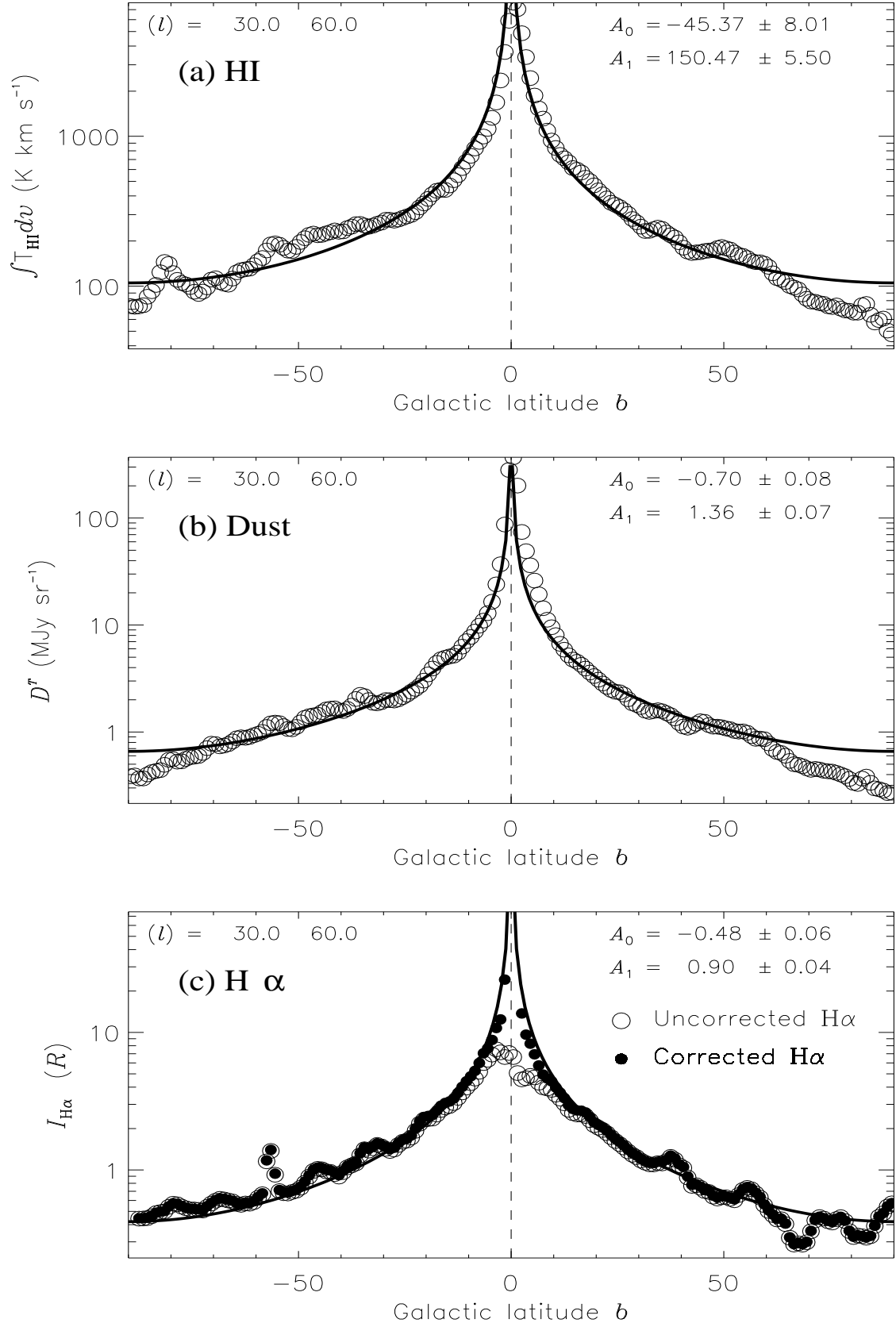}}
\end{picture}
\vskip 105mm
\caption[Galactic latitude scans for the longitude range
$l=30^{\circ}-60^{\circ}$.]{Galactic latitude scans for the longitude range
$l=30^{\circ}-60^{\circ}$. (a) H{\sc i} from Leiden-Dwingeloo survey data. (b) Dust
emissivity at 100 $\mu$m from the SFD98 $D^{T}$ template. (c) \ha data
from the WHAM survey. In (c) the \ha data corrected for dust absorption
($f_{d}=0.33$) are
shown as filled circles while the uncorrected data are open circles. The
best-fitting cosecant law is shown for each distribution (see text). Note the
logarithmic scale on the vertical axes.}
\label{galdep.fig}
 \end{figure*}

For neutral H{\sc i} gas we use the 1420 MHz Leiden-Dwingeloo northern sky
survey (Hartmann \& Burton 1997), \ha data
from the WHAM survey and the SFD98 100 $\mu$m $D^{T}$ map
to trace the dust. Each map was resampled onto a Galactic
coordinate grid and smoothed to a resolution of 1$^{\circ}$. Along
each 1$^{\circ}$-wide latitude strip, the values are averaged over a range of
longitudes to give a representative value for that latitude. Fig. \ref{galdep.fig} shows the latitude dependence of the gas and
dust in the region $l=30^{\circ}-60^{\circ}$. Both the H{\sc i} gas and $100~
\mu$m dust follow similar cosecant-law trends. Assuming a homogeneous slab of
material viewed at different latitudes $b$, the column density along a line of sight
is expected to follow a cosecant law when viewed from the central plane;
\begin{equation}
\label{cosecant.eqn}
~~~~~~~~~~~I = A_{0} + A_{1}/{\rm sin}(|b|)
\end{equation}
\noindent where $A_{0}$ is an offset and $A_{1}$ is the amplitude of the
cosecant law. This is indeed found to be the case, with deviations occurring due
to large concentrations of emitting material such as extended H{\sc ii}
regions and features like the Local System (Gould Belt). The region
$l=30^{\circ}-60^{\circ}$ is one of the ``cleaner'' regions of the
plane where no large structures exist. This region includes are own spiral arm and the Sagittarius
arm; here the dust emission on the Galactic ridge reaches an average of 500 MJy sr$^{-1}$ corresponding to $A$(\ha$\!\!$)$=
23$ mag for an extragalactic object. At intermediate and high
latitudes ($|b|>15^{\circ}$), $D^T$ is $\la 8$ MJy sr$^{-1}$ and the \ha absorption is
therefore small.

 The best-fitting cosecant law is a good match
to the H{\sc i} and dust distributions as shown in
Fig. \ref{galdep.fig}. This figure also shows
that the \ha intensity is greatly attenuated at low latitudes
($|b|<15^{\circ}$) due to the increased dust absorption nearer to the
plane. Note also that there is a significant offset $A_{0}$;
this corresponds to the Sun being in a local ``hole''. The $A_{0}/A_{1}$ values suggests that there is a
$\sim 30-50$ per cent deficit in the local slab density.
If one adopts equation (\ref{dust_absorption.eqn}) as the model for \ha absorption we can correct the \ha emission, adopting a
value of $f_{d}$ as discussed in Section \ref{abs_gal}. We can
then determine a value for $f_{d}$ by correcting the \ha latitude
distribution for absorption, so that it matches that of the gas and
dust distributions shown in Fig \ref{galdep.fig}. By varying $f_{d}$ we
find the best-fitting value is given by $f_{d}=0.33^{+0.10}_{-0.15}$. The
fit was made by deriving the cosecant law using data at $|b| >
20^{\circ}$ and fitting $f_{d}$ to this cosecant law over $|b| = 5^{\circ}$ to
$15$\degr. For $|b|<5^{\circ}$, the absorption is likely to be $> 1$
mag and therefore any derived value for $f_{d}$ becomes unreliable. The value $f_{d}=0.33$ derived here is representative of the solar
neighbourhood of our Galaxy within a few kilo-parsecs. Other longitudes
gave similar results, although the cosecant fitting is difficult due
to local structures such as the Gould Belt system and other bright
extended H{\sc ii} regions. Further information on the value of
$f_{d}$ is obtained from the study
of the radio free-free emission from several nearby H{\sc ii}
complexes given in Section \ref{obs_tests}.

The value of $f_{d}$ derived in this section is not consistent with
uniform mixing of ionized gas and dust which requires $f_{d} = 0.5$. However,
there is a large uncertainty in $f_{d}$ and it is expected that
$f_{d}$ will vary significantly across the sky. However, as a first-order
approach, we will
adopt the value $f_{d} = 0.33$ when estimating the free-free template
in Section \ref{free-free_template}. This lower value is indicative of non-uniform ionization; for example, a region of cloud is ionized by the interstellar radiation field which is in a narrower
Galactic layer of O,B stars.


\subsection{The \ha absorption template}
\label{extinction_template}

Using equation (\ref{dust_absorption.eqn}) we can convert the $D^{T}$
dust map to an absorption map which can be applied to the H$\alpha$
intensity $I_{{\rm H}\alpha}$, using

\begin{equation}
~~~~~~~~~~~I_{{\rm H}\alpha}^{\rm corrected} = I_{{\rm H}\alpha}
\times 10^{D^{T} \times 0.0185 \times f_{d}} 
\label{true_ha.eqn}
\end{equation}

\begin{figure*}
\setlength{\unitlength}{1mm}
\begin{picture}(80,92)
\put(0,0){\includegraphics{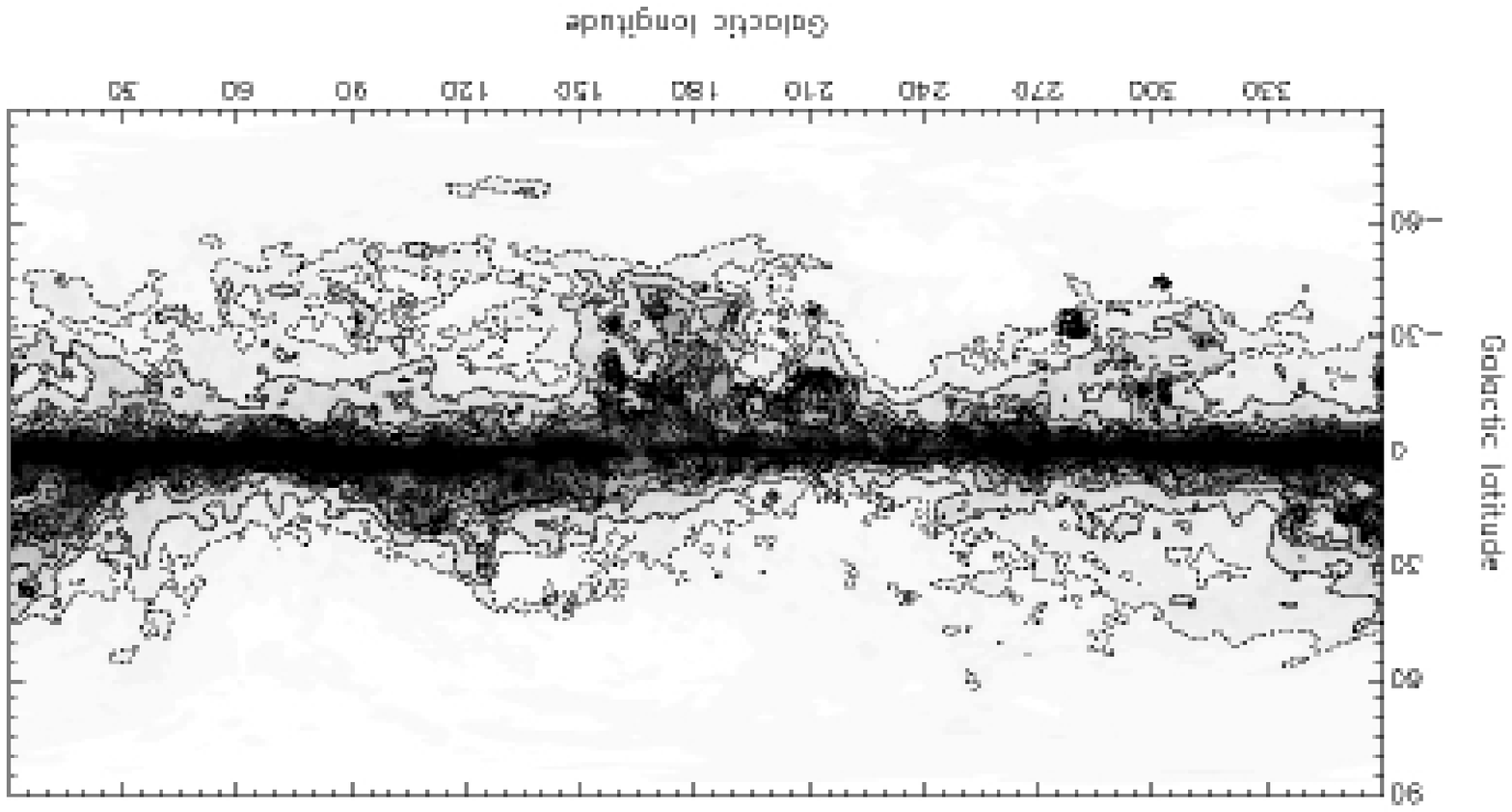}}
\end{picture}
\vskip 3mm
\caption[The \ha absorption template]{The absorption template to be
applied to the Galactic \ha emission. This assumes an absorbing
dust-fraction of $f_{d}=0.33$, as derived in Section
\ref{latitude_scans}, and the the total extragalactic value $A$(H$\alpha$) given in equation
(\ref{dust_absorption.eqn}). Linear grey-scale and contours show absorption
in magnitudes. Contours are given at 0.05
(dot-dashed),0.1,0.2,0.3,0.4,0.5 and 0.7
mag. Absorption above 1 mag is black on the grey-scale.}
\label{extinction_template.fig}
\end{figure*}

\noindent The
value of $f_{d}$ is taken to be 0.33. The absorption template to be
applied to Galactic \ha and smoothed to a resolution of $1^{\circ}$ is
shown in Fig. \ref{extinction_template.fig}.

The template clearly shows that the absorption of \ha by dust
is modest - less than 0.2 mag (20 per cent correction) for most of the
sky. Beyond $|b| \ga 50^{\circ}$, it is less than 0.05 mag (5 per
cent correction). At $|b| \la 5$\degr, the
absorption increases beyond 1 mag for much of the Galactic
plane. Individual dust clouds can have $ > 1$ mag of absorption at
intermediate latitudes. For a major portion of the sky the dust
correction clearly needs to be taken into account. For example, the correction of more than 50 per
cent is required in a limited area of the sky such as the Gould Belt system.

It is interesting to note that the absorption can be low ($< 0.5$ mag) close to the Galactic plane in the longitude range $160^{\circ}
< l < 260^{\circ}$. These regions may be useful in
testing the dust absorption and \ha templates at low
latitudes. However, for the remainder of the Galactic plane
($|b|<5^{\circ}$) the
absorption is too great to make any reasonable estimate of \ha emission. 

The uncertainty in deriving the \ha absorption from the dust using
equation (\ref{dust_absorption.eqn}) is considered to be 20 per cent
(see Section \ref{A_halpha}). A second factor in the uncertainty is
the relative distribution of gas (H$\alpha$) and dust in the line of
sight ($f_{d}$). Near the Galactic plane where there are many clouds/features in
the line of sight, our value of $f_{d}=0.33$ is realistic, while at higher latitudes where there are only a few clouds
in the line of sight, the uncertainty (in a smaller total absorption)
is greater. If we adopt a 20 per cent uncertainty in $f_{d}$ and
combine it with a 20 per cent error in equation
(\ref{dust_absorption.eqn}), we obtain a total uncertainty of 30 per
cent in the correction to the \ha intensity for dust absorption. Thus if there is an \ha absorption ($f_{d} \times A$(H$\alpha$)) of 1.0 mag, the 30 per
cent uncertainty in multiplying factor required to give the estimate
of the true \ha intensity is $2.5\pm0.7$. However for a total
absorption of 0.2 mag the multiplying factor is $1.20 \pm 0.07$. This
illustrates the higher fractional accuracy of estimating the \ha
intensity at lower levels of absorption. In summary, the absorption of
\ha can be estimated up to a limit of $\sim 1$ mag; this
corresponds to $D^{T}=65$ MJy sr$^{-1}$ for $f_{d}=0.33$ and is shown
as black on Fig. \ref{extinction_template.fig}.


\section{Conversion of \ha to free-free continuum emission}
\label{halpha_conversion}

The ionized interstellar medium generates both radio free-free
continuum and \ha emission. Ionized hydrogen alone is
responsible for the \ha emission while ionized hydrogen and
helium produce the radio continuum. Both the optical and radio
emission are functions of electron temperature, $T_{e}$. A relationship between
the two can be determined on certain assumptions. We will derive this
relationship and give an estimate of the uncertainties which come
from the theory and from the variation in the electron temperature
at different locations within the Galaxy.

\subsection{\ha emission}
\label{ha_emission}

\begin{figure}
\setlength{\unitlength}{1mm}
\begin{picture}(80,55)
\put(0,0){\includegraphics{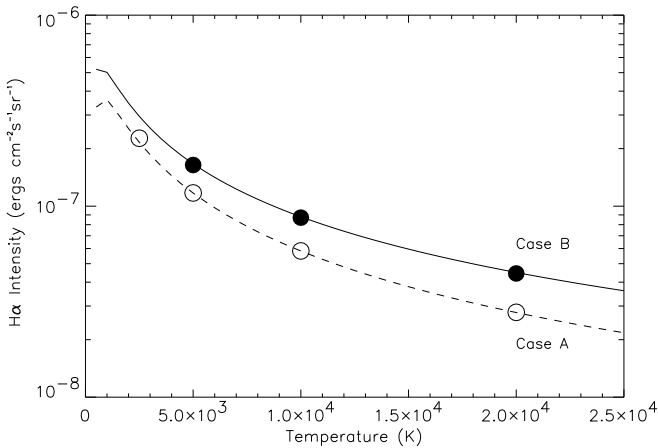}}
\end{picture}
\vskip 3mm
\caption[\ha emission as a function of $T_{e}$ for case A and
case B]{\ha emission per unit Emission Measure as a function of $T_{e}$ for case A (optically
thin to Lyman photons) and case B (optically thick to Lyman
photons) for $n_{e} = 10^{2}$ cm$^{-3}$. The circles represent the values calculated by theory given
by Hummer \& Storey (1987). The curves are the expressions given by Valls-Gabaud (1998); case B is given in equation (\ref{vg_b.eqn}).}
\label{caseA-B.fig}
 \end{figure}

The Balmer line emission from ionized interstellar gas is well
understood; a clear discussion is given by Osterbrock
(1989). However, the \ha line intensity depends upon whether the emitting medium is optically thick (case B)
or optically thin (case A) to the ionizing Lyman continuum. In the
H{\sc ii} regions and nebulae studied using \ha it is believed
that case B applies (Osterbrock 1989). Knowing the emission coefficients for
both case A and case B for H$\beta$ and using the Balmer decrements
(Hummer \& Storey 1987), the \ha emission can
be readily calculated for the temperature range $5000-20000$ K both for case A
and case B. Fig. \ref{caseA-B.fig} shows the \ha intensity per unit
Emission Measure ($EM \equiv \int n_{e}^2 dl$) for
$T_{e}=5000,10000,20000$ K for both case A and case B.

The emission coefficients are generally accepted to be accurate
to $\approx 1$ per cent (Pengelly 1964; Hummer \& Storey 1987). For case B,
the intensity varies very weakly with number density, amounting to a
few percent over 2 orders of magnitude in density. The values
shown in Fig. \ref{caseA-B.fig} are for a density ($n_{e} = 10^2$
cm$^{-3}$) gas. Case A gives $\approx 30$ per cent lower intensity
than case B over the likely temperature range. 

The
situation for intermediate and high latitudes appears to favour case B
on the following arguments. For case B to apply, the optical depth $\tau$, to
Lyman continuum photons ($\lambda \ltsim 900$ \AA) must be greater
than unity in the emitting region. Pottasch (1983) gives the mean free
path of Lyman continuum photons as $\approx 0.052 n_{\rm H}^{-1}$ pc,
corresponding to an optical depth relation $\tau = 19.2 n_{\rm H}l$
where $l$ is path length in pc. For a typical H{\sc ii} region in the
Galactic plane $n_{e} \approx n_{\rm H} = 10^{2} - 10^{6}$
cm$^{-3}$ on typical scales of $1-10$ pc and case B conditions are
clearly satisfied with $\tau \approx 10^{3}-10^{10}$. At intermediate
latitudes faint \ha features with $EM \sim 1$
cm$^{-6}$ pc can be identified with scale sizes of $\sim 10-100$
pc. The corresponding $n_{e}$ is $0.1-0.3$ cm$^{-3}$. The optical depth to
Lyman continuum in these low density regions is then $1-30$, still
satisfying the case B conditions.

A useful and accurate expression for the \ha
intensity based on data from Hummer \& Storey (1987) is given by
Valls-Gabaud (1998) for case B in units of ${\rm erg~ cm^{-2}~ s^{-1}~ sr^{-1}}$:

\begin{equation}
\label{vg_b.eqn}
I({\rm H} \alpha) \stackrel{{\rm Case B}}{=} 9.41 \times 10^{-8} T_{4}^{-1.017}
10^{-0.029/T_{4}} \times (EM)_{{\rm cm^{-6}pc}}
\end{equation}
where $T_{4}$ is the electron temperature in units of $10^{4}$ K. This is plotted as a continuous curve along with the equivalent curve
for case A in Fig. \ref{caseA-B.fig}, which shows that the agreement of equation
(\ref{vg_b.eqn}) with theory is better than 1
per cent over the temperature range $5000-20000$ K. 

\subsection{Radio continuum emission}
\label{radio_cont}

The free-free radio continuum emission for an ionized gas at $T_{e} <
550,000$ K and in Local
Thermal Equilibrium (LTE) is well described in terms of a volume
emissivity in units of ${\rm erg}~{\rm cm^{-3} s^{-1}
Hz^{-1}}$ (Oster 1961):

\begin{equation}
\epsilon_{\nu}^{ff} = 6.82 \times 10^{-38} Z^{2} n_{{\rm e}} n_{{\rm ion}}
T_{e}^{-1/2}e^{-h \nu / kT} <\bar{g_{ff}}> 
\end{equation} 

\noindent Here $<\bar g_{ff}(Z, \nu, T_{e})>$ is the velocity-averaged Gaunt
factor. The Gaunt factor comprises all the terms by which the quantum
mechanical expressions differ from the classical ones. The evaluation of the Gaunt factor  has been
refined over recent years since the work of Scheuer (1960) and Oster
(1961) for a range of frequencies, temperatures. Karzas \& Latter (1961) have given definitive expressions for $\bar{g}_{ff}$. Hummer (1988) has used
these expressions to obtain an accurate analytical expansion for
$\bar{g}_{ff}$ in terms of a two-dimensional Chebyshev fit covering a
wide range of frequencies and temperatures. Table
\ref{gaunt_factors.table} lists accurate Gaunt factors (accurate to 0.7 per cent) derived from
Hummer (1988) for a range of frequencies and temperatures relevant to
CMB studies. 

\begin{center}
\begin{table}
\centering
\caption{Gaunt factors for relevant frequencies and electron
temperatures calculated from Hummer (1988) and accurate to 0.7 per cent.}
\begin{tabular}{||l||c|c|c|c|c||}
\hline
$\nu$ (GHz) & &$T_{e}$ &(K) & \\ 
 & 4000 & 6000 &7000 &8000 & 10000 \\ \hline \hline
0.4 &5.82 &6.03 &6.11   &6.18 &6.39 \\ \hline 
1.4 &5.16 &5.37 &5.45   &5.52 &5.73 \\ \hline
2.3 &4.90 &5.11 &5.19   &5.26 &5.47 \\ \hline
10 &4.13 &4.34 &4.42 &4.49 &4.70   \\ \hline
30 &3.55 &3.76 &3.84    &3.91 &4.12 \\ \hline
44 &3.35 &3.56 &3.64    &3.71 &3.92 \\ \hline
70 &3.10 &3.32 &3.40    &3.45 &3.67 \\ \hline
100 &2.92 &3.13 &3.21   &3.28 &3.49 \\ \hline
\label{gaunt_factors.table}
\end{tabular}
\end{table}
\end{center}

\begin{figure}
\setlength{\unitlength}{1mm}
\begin{picture}(80,55)
\put(0,0){\includegraphics{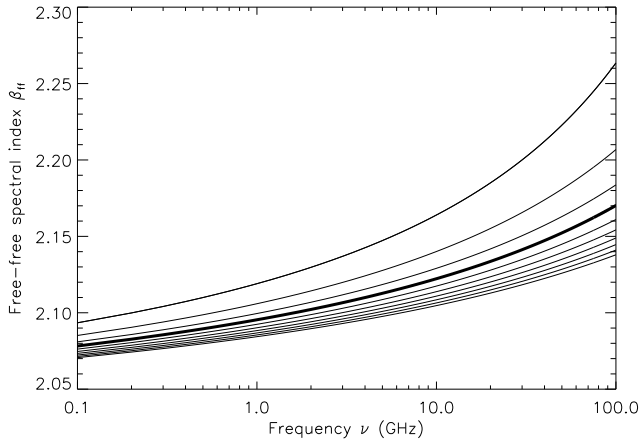}}
\end{picture}
\vskip 3mm
\caption[Free-free spectral index]{The local free-free spectral index
as a function of frequency and electron temperature; given by
equation (3) of Bennett et al. (1992). The curves from top to bottom are for temperatures
$T_{e}=$(2,4,6,8[heavy],10,12,14,16,18,20)$\times 10^{3}$ K.}
\label{beta_ff.fig}
 \end{figure}

\noindent We now summarize the present situation for the frequencies
($0.4-100$ GHz) covered by the observations analysed here and the
observing range of the Low Frequency Instrument (LFI) on the {\it Planck} Surveyor satellite. The free-free spectral index
$\beta_{ff}$ is a slow function of frequency and electron temperature
(Bennett et al. 1992) as shown in Fig. \ref{beta_ff.fig}. Over a wide range in frequencies
this can lead to a significant discrepancy in the predicted radio
emission if this is not accounted for. One should therefore still use
the accurate formalism given by Oster
(1961) who derives the optical depth for free-free emission as

\begin{eqnarray}
\tau_{c} ({\rm Oster}) = 3.014\times 10^{-2} T_{e}^{-1.5} {\nu_{{\rm GHz}}^{-2.0}}
\times \nonumber \\
\left \{ {\rm ln}[4.955 \times 10^{-2} \nu^{-1}] + 1.5{\rm ln}(T_{e})
\right \} \times (EM)_{{\rm cm^{-6}pc}}
\end{eqnarray}

\noindent The approximation given by Altenhoff et al. (1960) is often used:

\begin{equation}
\tau_{c} ({\rm AMWW}) = 8.235 \times 10^{-2} T_{e}^{-1.35} \nu_{{\rm
GHz}}^{-2.1} \times (EM)_{{\rm cm^{-6}pc}}
\end{equation}

\noindent For low frequencies ($\la 1$ GHz) the error in this expression is of the order a few percent
but increases to 5-20 per cent for frequencies above 10 GHz. Mezger \& Henderson (1967) define a factor $a(T,\nu)$, the ratio
between the two formulae as 

\begin{eqnarray}
\label{a_factor.eqn}
a = \frac{\tau_{c} ({\rm Oster})}{\tau_{c}({\rm AMWW})} = 0.366
\nu_{{\rm GHz}}^{0.1}
T_{e}^{-0.15} \times \nonumber \\
\left \{ {\rm ln}[4.995 \times 10^{-2} \nu_{{\rm GHz}}^{-1}] +
1.5{\rm ln}(T_{e}) \right \}
\label{a_factor.eqn}
\end{eqnarray}

\begin{center}
\begin{table}
\centering
\caption[The $a$ factors for relevant frequencies and
temperatures]{The factor $a = \tau_{c} ({\rm Oster}) / \tau_{c} ({\rm AMWW})$ for relevant frequencies and temperatures using equation (\ref{a_factor.eqn}).}
\begin{tabular}{||l||c|c|c|c|c||}
\hline
$\nu$ (GHz) & &$T_{e}$ &(K) & \\ 
 & 4000 & 6000 &7000 &8000 & 10000 \\ \hline \hline
0.4 &0.9972 &0.9934 &0.9912 &0.9889 &0.9844 \\ \hline 
1.4 &0.9936 &0.9974 &0.9978 &0.9977 &0.9967 \\ \hline
2.3 &0.9872 &0.9946 &0.9962 &0.9972 &0.9978 \\ \hline
10 &0.9484 &0.9684 &0.9745 &0.9792 &0.9857 \\ \hline
30 &0.8957 &0.9276 &0.9379 &0.9461 &0.9582 \\ \hline
44 &0.8717 &0.9084 &0.9203 &0.9299 &0.9442 \\ \hline
70 &0.8382 &0.8810 &0.8952 &0.9066 &0.9238 \\ \hline
100 &0.8090 &0.8569 &0.8729 &0.8858 &0.9054 \\ \hline
\label{a_factor.table}
\end{tabular}
\end{table}
\end{center}

Table \ref{a_factor.table} gives accurate values of $a$ based on
equation (\ref{a_factor.eqn}) for a range of temperatures
and frequencies. The $a$ factor is convenient in that it allows a
simple formula to be written down using a single spectral index and
temperature dependence which are modified by the Gaunt factor. 

The brightness temperature $T_{b}$ is given by

\begin{equation}
T_{b} = T_{e} (1 - e^{- \tau_{c}}) \approx T_{e} \tau_{c} ~~{\rm for}~
\tau_{c} << 1
\end{equation}

\noindent For the frequencies relevant to CMB observations and the diffuse
interstellar medium discussed here, the optically thin assumption
is valid. At 1 GHz only the brightest H{\sc ii} regions in the Galaxy
are optically thick ($EM \ge 10^{6}$ cm$^{-6}$ pc) while the Galactic
ridge has $\tau \sim 10^{-3}$ ($EM \ge 10^{3}$ cm$^{-6}$ pc
$\approx 500~R$) estimated for $T_{e}=7000$ K.

Finally, the brightness temperature in Kelvin can now be written as
\begin{equation}
\label{tb.eqn}
T_{b} = 8.235 \times 10^{-2} a  T_{e}^{-0.35} \nu_{{\rm GHz}}^{-2.1} (1 +
0.08) \times (EM)_{{\rm cm^{-6}pc}}
\end{equation}
where the factor (1+0.08) is the contribution from the fraction of He atoms, all of which
are assumed to be singly ionized; $a$ is given by equation (\ref{a_factor.eqn}). 


\subsection{Electron temperatures in Galactic H{\sc ii} regions}
\label{temps}

Optical determinations of electron temperatures in H{\sc ii} regions
using forbidden line ratios give values in the range 5000 to 20000 K
with a mean value of $\approx 8000$ K (Reynolds 1985). These
determinations refer to the solar neighbourhood within 1 or 2 kpc of the Sun.

Electron temperature determinations using extensive radio
recombination line (RRL) data probe the greater part of the Galaxy and
show a clear temperature gradient as a function of galacto-centric
radius. The results of Shaver et al. (1983) for 67 H{\sc ii}
regions show a clear gradient in
temperature increasing with galacto-centric radius from 5000 K at 4 kpc to 9,000 K at 12
kpc. This is due to the decreasing metallicity at larger
radii (Panagia 1979). There is a spread of approximately 1000 K at a particular
radius. The local value at $R_0 = 8.5$ kpc is $T_{e}=7000$ K and will
be adopted as the typical value in the absence of other information.

\subsection{Expected radio emission from \ha maps}
\label{expected}

\begin{figure}
\setlength{\unitlength}{1mm}
\begin{picture}(80,55)
\put(0,0){\includegraphics{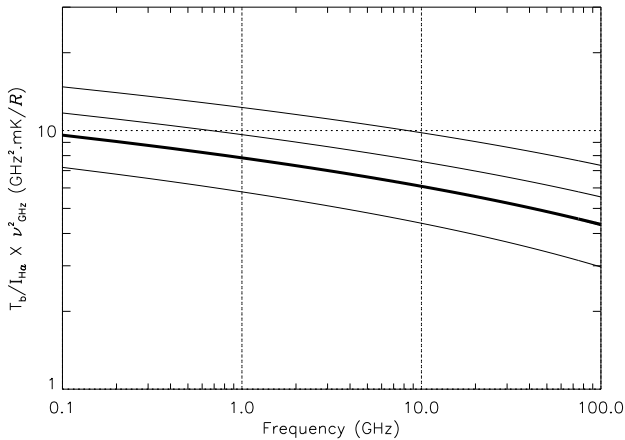}}
\end{picture}
\vskip 3mm
\caption[Ratio of radio to \ha]{Ratio of radio brightness temperature
to \ha intensity multiplied by $\nu^{2}_{\rm GHz}$ as a function of
frequency. Curves are for $T=$4000,7000[heavy],10000 and 15000K which
is the likely temperature range corresponding to a factor $\sim 2.4$ in
the ratio. The curvature in each curve reflects the frequency
dependence of the Gaunt factor.}
\label{ratio.fig}
\end{figure}

The relationship between radio emission and \ha emission can now
be calculated using equations (\ref{vg_b.eqn}) and (\ref{tb.eqn}), in units of mK/$R$ as 
\begin{equation}
\label{tb_ha.eqn}
\frac{T_{b}^{ff}}{I_{{\rm H}\alpha}} = 8.396 \times 10^{3} a \times
\nu_{{\rm GHz}}^{-2.1} T_{4}^{0.667} 10^{0.029/T_{4}} (1+0.08)
\end{equation}
where $T_{4}$ is the electron temperature in units of $10^4$ K. The conversion factors for relevant frequencies are given in Table
\ref{halpha_conversions.table} for $T_{e}=7000$ K. Fig. \ref{ratio.fig}
shows $(T_{b}^{ff}/I_{{\rm H}\alpha}) \nu^{2}_{\rm GHz}$ over the
frequency range 100 MHz to 100 GHz for $T_{e} = 4000,7000,10000$ and
$15000$ K. The increase in (negative) slope with frequency is due to
the effect of $a$ arising from the Gaunt factor dependence. The
variation in $T_{b}^{ff}$ at a given frequency corresponds to a factor of $\sim 2.4$ between 4000 and
15000 K. 

It is of interest to compare our equation (\ref{tb_ha.eqn}) with
recent expressions given in the literature. Reynolds \& Haffner (2002)
give $T_{b}^{ff}/I_{H\alpha} = 7.4 \nu_{30}^{-2.14}~\mu {\rm K}/R$ at
$T_{e} = 8000$ K. Our value at 30 GHz and 8000 K would be 17 per cent
lower at 6.35 $\mu$K$/R$. This difference may be attributed to the 17 per
cent over-estimate of the Gaunt factor in equation (6) of Valls-Gabaud
(1998) which are used by Reynolds \& Haffner (2002). We have resolved
this difference with help from David Valls-Gabaud (private communication). The coefficient in
his equation (6) should be 3.96 and not 4.4; furthermore, a
typographical error in his equation (11) gave the exponent of $T_{4}$
as 0.317 instead of 0.517. He agrees with the calculations of the
Gaunt factors in Section \ref{radio_cont}. His new calculations also agree
well with our equation (\ref{tb_ha.eqn}). Equation (\ref{vg_b.eqn}), also from
Valls-Gabaud (1998), is not affected.

The uncertainty in predicting the radio free-free emission from the
\ha emission as given in equation (\ref{tb_ha.eqn}) arises mainly from
the uncertainty in the electron temperature. The $\pm 1000$ K
uncertainty in the adopted solar neighbourhood value of 7000 K is an
upper limit to the scatter of H{\sc ii} region temperatures since it
includes some measurement errors. The corresponding error in
equation (\ref{tb_ha.eqn}) is 10 per cent. The other possible
uncertainty is in whether case A or case B applies in the lower
density regions at intermediate and high Galactic latitudes of
interest to CMB studies. For the reasons given in Section \ref{ha_emission}, we will adopt the generally assumed position
that case B is correct and take the associated error in equation
(\ref{vg_b.eqn}) to be 1 per cent. However, in the unlikely event that
case A were to apply, the value of $T_{b}/I_{{\rm H}\alpha}$ given in
equation (\ref{tb_ha.eqn}) would be increased by 30 per
cent. Uncertainties in equation (\ref{vg_b.eqn}) and in the ionized helium to
hydrogen ratio in equation (\ref{tb.eqn}) are only a few per cent. We
consider that the overall uncertainty in applying equation (\ref{tb_ha.eqn}) to
be $\sim 10$ per cent for a large sample of the sky, assuming case B
applies. 

\begin{center}
\begin{table}
\centering
\caption{Free-free brightness temperatures per unit Rayleigh for a
range of frequencies assuming $T_{e} = 7000$ K.}
\label{halpha_conversions.table}
\begin{tabular}{||c||c||}
\hline
Frequency &Free-free brightness temperature \\ 
$[{\rm GHz}]$ &per unit Rayleigh \\ \hline \hline
0.408 & 51.2 mK  \\ \hline
1.420 & 3.76 mK \\ \hline
2.326 & 1.33 mK \\ \hline
10 &60.9 $\mu$K \\ \hline
30  &5.83 $\mu$K \\ \hline
44  &2.56 $\mu$K \\ \hline
70  &0.94 $\mu$K \\ \hline 
100 &0.43 $\mu$K \\ \hline
\label{conversions.table}
\end{tabular}
\end{table}
\end{center}


\section{Observational tests of the \ha to free-free relation}
\label{obs_tests}

We now make an observational test of equation (\ref{tb_ha.eqn}) which predicts the free-free
radio emission for a given \ha intensity. In practice, the diffuse
free-free emission is mixed with synchrotron emission so the free-free
component must be identified and separated. The free-free emission can
be identified in two ways. Firstly, by its associated radio
recombination line emission and secondly, by its free-free spectrum
using multi-frequency observations. Furthermore, a comparison between the radio and
\ha will lead to an estimate of the \ha absorption by dust; this can be compared with $A(H\alpha)$ given by equation
(\ref{dust_absorption.eqn}) to give a value of the absorbing factor of
dust $f_{d}$ in front of the \ha$\!\!$.

\subsection{Radio Recombination Lines}
\label{RRL}

One advantage of using radio recombination lines (RRL's) is that the
ratio of the line temperature to continuum temperature can be used to
derive an electron temperature $T_{e}$ for the H{\sc ii} emission
which can be substituted in equation (\ref{tb_ha.eqn}) to give the
radio emission expected from the observed \ha emission. Two extended
regions of \ha emission which are sufficiently bright to have been
studied in RRL's are Barnard's Arc and the Gum Nebula.

\subsubsection{Barnard's Arc}
\label{barnards_arc}

Barnard's Arc is an ionization-bounded H{\sc ii} region at 460 pc
photo-ionized by the Orion I OB star association (Reynolds \& Ogden 1979). At radio frequencies Barnard's Arc has a thermal spectral index
(Davies 1963). Gaylard (1984) describes RRL observations at 2.272 GHz
at 3 positions in the Arc. The radio data were converted from antenna
to brightness temperatures using the factor 1.5 used by Gaylard
(1984). The \ha observations from SHASSA and the
$D^{T}$ dust map were smoothed to 20 arcmin resolution for this
study. Table \ref{RRL.table} gives the observed brightness temperature
$T_{b}$, $T_{e}$ and $T_{{\rm H}\alpha}$, the brightness temperature
expected from the observed \ha intensity $I_{{\rm H}\alpha}$, uncorrected for
absorption using $T_{e}$ values also given in Table \ref{RRL.table}. The small ($<10$ per cent) errors in
the radio data for the 3 positions in Barnard's arc, indicate that the
calculated $f_{d}$ values are accurate to
about $\pm~0.15$ and are therefore consistent with no or little absorption
($f_{d} \sim 0$) in all 3 positions. No absorption would be expected if the \ha$\!\!$-emitting gas were nearer than the dust, in
this nearby ($<500$ pc) Gould Belt feature.
\begin{center}
\begin{table*}
\centering
\caption[Comparison of RRL data and \ha predictions]{Comparison of the observed brightness temperature $T_{b}$
from RRL data with the predicted brightness temperature $T_{{\rm H}\alpha}$ from
\ha data using equation (\ref{tb_ha.eqn}). The first three positions are in
Barnard's Arc (Gaylard 1984) and the next three are in the Gum Nebula
(Woermann et al. 2000). Using the dust column density $D^{T}$ and
electron temperature $T_{e}$, the fraction of $f_{d}$ dust causing absorption
of \ha emission is calculated for each observation.}
\begin{tabular}{||l|l|c|c|c|c|c|c|c||}
\hline
Coordinates &Frequency &$T_{b}$ &$T_{e}$ &$I_{{\rm H}\alpha}$ &$D^{T}_{100 \mu m}$
&$T_{{\rm H}\alpha}$ &$f_{d}$ \\
(B1950/Galactic) &[MHz] &[mK] &[K] &[R] &[MJy sr$^{-1}$] &[mK] & \\ \hline \hline
$5^{h}49^{m} +00^{\circ}38^{m}$ &2272.661 &$295 \pm 24$ &$5200 \pm 400$
&$280 \pm 24$ &21.2
&330 &$-0.1$ \\ \hline
$5^{h}51^{m} -01^{\circ}00^{m}$ &2272.661 &$280 \pm 18$ &$3800 \pm 300$
&$170 \pm 15$ &15.2
&269 &+0.1 \\ \hline
$5^{h}54^{m} -02^{\circ}45^{m}$ &2272.661 &$258 \pm 14$ &$6600 \pm 1100$
&$188 \pm 17$ &17.2
&254 &0 \\ \hline \hline
G264.00 $-12.80$ &1715.67 &$450 \pm 105$ &$8500 \pm 2400$ &$223 \pm 20$ &6.7 &673 &$-1.4$ \\
 &2422.46        &$263 \pm 53$ &$7800 \pm 2300$&$223 \pm 20$ &6.7 &310 &$-0.6$ \\ \hline
G265.00 $-08.00$ &1715.67 &$570 \pm 120$ &$5700 \pm 1400$ &$232 \pm 21$ &12.0 &535 &+0.1 \\ \hline
G270.65 +10.40 &1715.67 &$398 \pm 90$ &$5800 \pm 1400$ &$123 \pm 11$ &6.2 &302
&+1.0 \\ \hline \hline
\label{RRL.table}
\end{tabular}
\end{table*}
\end{center}
\subsubsection{The Gum Nebula}

The Gum Nebula is a diffuse emission region centred on $(l,b)=
(258^{\circ},-2^{\circ})$ (Gum 1952). It is a complex at a distance of 500 pc
containing both free-free and synchrotron features which are probably
the consequence of supernova activity (Duncan et al. 1996; Reynosa \&
Dubner 1997). The RRL data from Woermann et al. (2000) for \ha features within the complex taken with 20 arcmin resolution are
compared with the \ha and dust data as in Section \ref{barnards_arc}
and are given in Table \ref{RRL.table}. The data tabulated are for the
3 strongest \ha features in the Gum Nebula. There is a large spread in
$f_{d}$ values from $-1.4$ to $+1.0$. The large error bars on
the radio data ($T_{b}, T_{e}$) correspond to an error of
approximately $\pm~0.6$ to $\pm~1.0$ in the derived $f_{d}$ values for
the three Gum nebula positions. Little can
be deduced about the value of $f_{d}$ without better radio data; negative
values have no physical meaning except that the \ha prediction is
larger than is measured in the radio.

\subsection{Multi-frequency scans across \ha features}
\label{scans}

\begin{figure*}
\setlength{\unitlength}{1mm}
\begin{picture}(80,135)
\put(0,0){\includegraphics{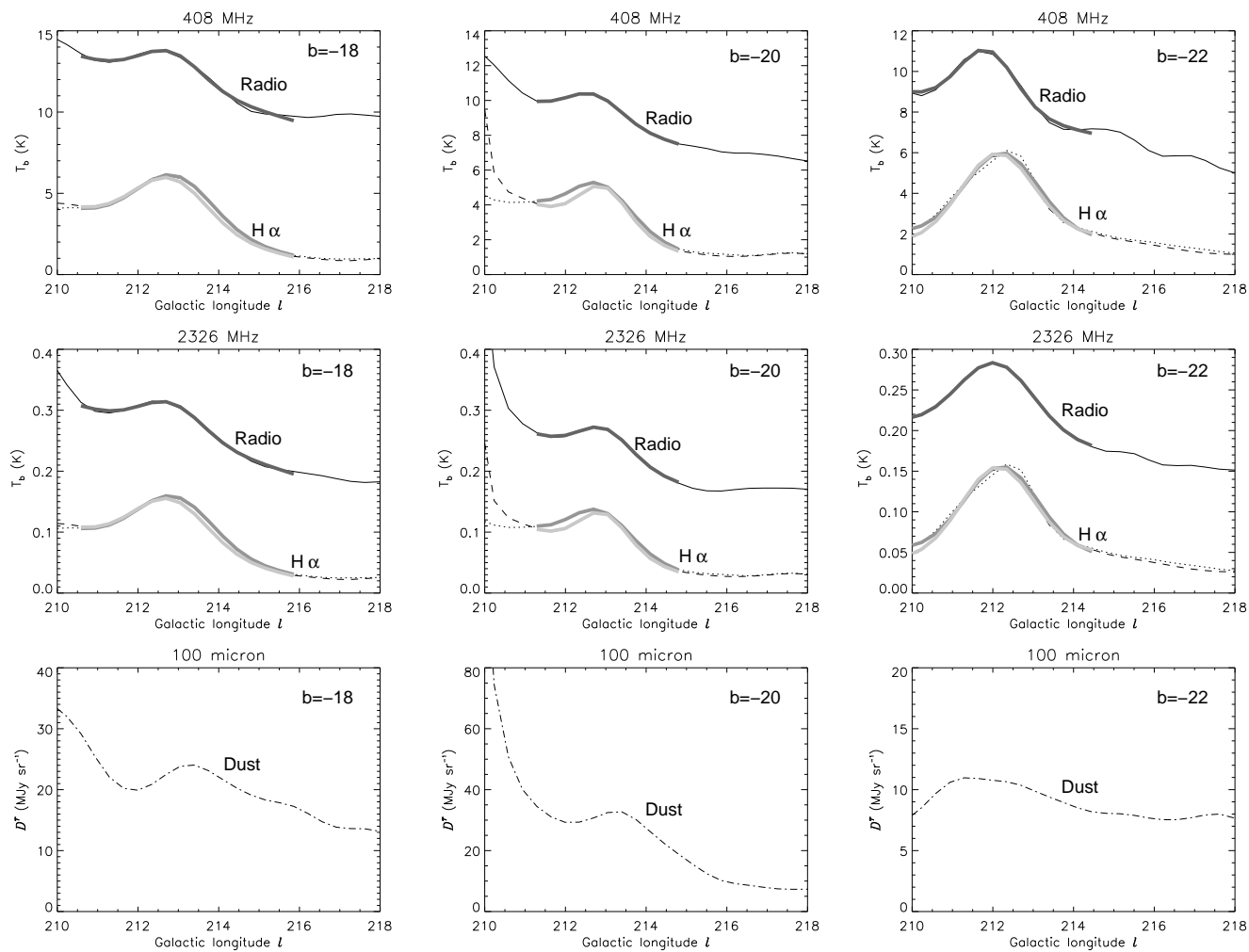}}
\end{picture}
\vskip 3mm
\caption[408 and 2326 MHz scans across Barnard's Arc]{408 and 2326 MHz
scans across Barnard's Arc at constant $b =
-18^{\circ}$ (left),$-20^{\circ}$ (middle) and $-22^{\circ}$
(right). The top panels show the 408 MHz scan and the prediction from
\ha$\!\!$; both WHAM and SHASSA are shown for each. WHAM is the dotted
line and SHASSA is the dashed line. The best-fitting gaussian plus a
linear baseline is shown as a heavy line; WHAM is the heavier line
and SHASSA is the lighter line. The middle panels show 2326 MHz data
and the same predictions as above for \ha$\!\!$. The bottom panel is the scan
for $D^{T}$ (SFD98).}
\label{BA_scans.fig}
 \end{figure*}

An alternative method of identifying and measuring the radio free-free
emission from an \ha feature is to use multi-frequency scans across it
to separate the free-free from the underlying synchrotron
emission. The brightest features of Barnard's Arc are ideal since they
lie some $20^{\circ}$ from the Galactic plane where the background synchrotron
emission is weak and relatively smooth. We have used the 408 MHz map
of Haslam et al. (1982) in the destriped form given by Davies et
al. (1996) and the 2326 MHz map of Jonas et al. (1998) to make 3
Galactic longitude scans across the Barnard Arc feature at $b =
-18^{\circ},-20^{\circ}$ and $-22^{\circ}$ centred on $l \sim 212$\degr. These
are illustrated in Fig. \ref{BA_scans.fig} along with the
corresponding \ha scans from WHAM data; all data are at a resolution
of 1\degr. The amplitudes of the best-fitting gaussians to the feature
centred on each \ha position are given in Table \ref{scans.table}.
Values of $f_{d}$ are estimated as in Section \ref{RRL} for $T_{e} =
7000$ K (the average local value) and $T_{e}=5100$ K (the weighted
average of $T_{e}$ from Table \ref{RRL.table}). The derived $f_{d}$ values are
accurate to about $\pm~0.1$ to $\pm 0.2$ and are significantly
negative when $T_{e}=7000$ K is assumed, indicating the radio emission is greater than expected from
the \ha brightness with no foreground absorption. However, when the
average measured value of $T_{e}$ (5100 K) is used, the derived
$f_{d}$ values are consistent with no absorption
($f_{d}=0$). This clearly shows the importance of having accurate
electron temperatures. In this case, the lower electron temperature
corresponds to a $\sim 25$ per cent reduction in $T_{{\rm H}\alpha}$. A
further consideration is the full-beam to main-beam correction (see Sections 6
\& 7) which can lead
to a factor of $\sim 1.5$ for objects comparable in size to the
beam. For Barnard's Arc, which is extended ($\sim 5^{\circ}$) in one
direction and $\sim 1^{\circ}\!.5$ in the perpendicular direction, the
correction is likely to be the square root of this factor ($\sim 1.2$). This further
correction would lead to a slightly positive ($\sim 0.2$) value for $f_{d}$.
\begin{center}
\begin{table*}
\centering
\caption[Comparison of multi-frequency radio data and \ha]{Comparison
of the observed radio brightness temperature $T_{b}$ with the
prediction from ${\rm H}\alpha$ data, $T_{{\rm H}\alpha}$, for 3 scans across Barnard's
Arc at frequencies of 408 and 2326 MHz. The \ha intensity $I_{{\rm
H}\alpha}$ and dust
column density $D^{T}$ are also listed. The expected brightness
temperature $T_{{\rm H}\alpha}$ and the associated dust fraction
actually absorbing $f_{d}$ are calculated for $T_{e}=7000$ K (columns
7 \& 8), and for $T_{e}=5100$ K (columns 9 \& 10) (see text).}
\begin{tabular}{||l|l|c|c|c|c|c|c|c||}
\hline
Coordinates &Frequency &$T_{b}$ &$I_{{\rm H}\alpha}$ &$D^{T}_{100 \mu m}$
&$T_{{\rm H}\alpha}$ [mK] &$f_{d}$ &$T_{{\rm H}\alpha}$ [mK] &$f_{d}$ \\
$(l,b)$ &[MHz] &[mK] &[R] &[MJy sr$^{-1}$] &($T_{e}=7000$ K) &($T_{e}=7000$ K)
&($T_{e}=5100$ K) &($T_{e}=5100$ K) \\ \hline \hline
$(212.9^{\circ},-18^{\circ})$ &408 &1980 &$67 \pm 7$ &22.5 &$3400
\pm 300$ &$-0.6$ &$2470 \pm 220$ &$-0.2$ \\
 &2326 &53 &$67 \pm 7$  &22.5 &$89 \pm 8$ &$-0.5$ &$64 \pm 6$ & $-0.2$ \\ \hline
$(212.9^{\circ},-20^{\circ})$ &408 &1490 &$47 \pm 5$ &30.7 &$2400 \pm 220$ &$-0.4$ &$1730 \pm 160$ &$-0.1$ \\
 &2326 &53 &$47 \pm 5$ &30.7 &$62 \pm 6$  &$-0.1$ &$45 \pm 4$ &$+0.1$ \\ \hline
$(212.0^{\circ},-22^{\circ})$ &408 &2990 &$80 \pm 8$ &10.8 &$4100 \pm 370$ &$-0.7$ &$2950 \pm 270$ &$0$ \\
&2326 &89 &$80 \pm 8$ &10.8 &$107 \pm 10$ &$-0.4$ &$85 \pm 8$ & $+0.1$
\\ \hline \hline
\label{scans.table}
\end{tabular}
\end{table*}
\end{center}

\subsection{Identifying \ha morphology in radio maps}
\label{subtract}

\begin{figure*}
\setlength{\unitlength}{1mm}
\begin{picture}(80,147)
\put(0,0){\includegraphics{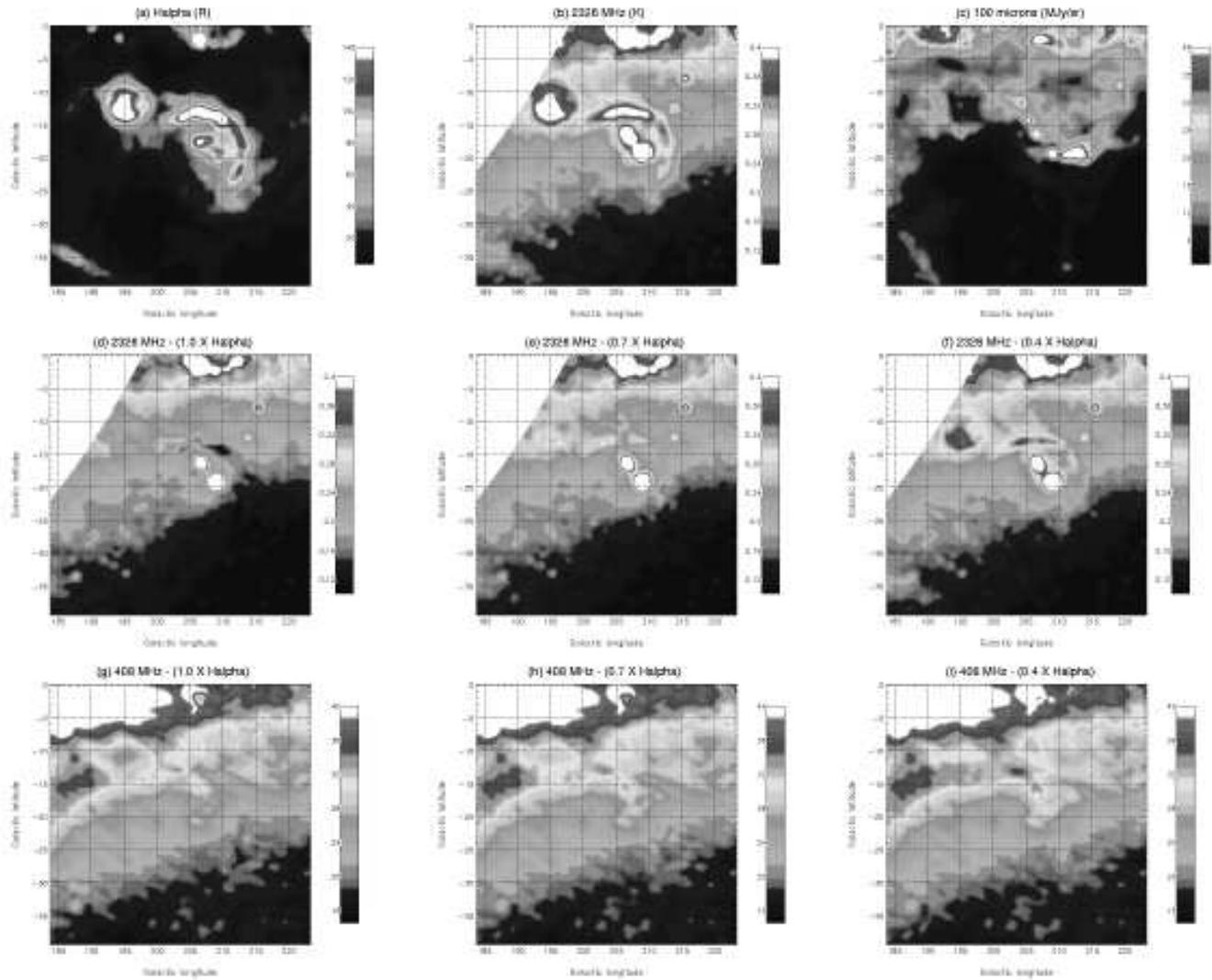}}
\end{picture}
\vskip 3mm
\caption[Morphology of the free-free component in the Orion
region]{Morphology of the free-free component in the Orion region. (a)
\ha data from WHAM clearly shows Barnard's Arc and
$\lambda$-Orionis. (b) 2326 MHz data from Jonas et al. (1998) showing a
similar morphology. (c) 100 $\mu$m $D^{T}$ map from SFD98 shows some correlation but not a
one-to-one correlation. Panels (d,e,f) show the 2326 MHz
map with varying levels (1.0,0.7 and 0.4 multiplied by the conversions
given in Table \ref{conversions.table}) of free-free emission subtracted as predicted
by the \ha data. Panels (g,h,i) are the same for the 408 MHz map from
Haslam et al. (1982). The electron temperature is assumed to be
$T_{e}=7000$ K. The best subtraction occurs when the factor is
$\sim 0.7$. Note that no correction has been made for dust
in this illustration and that the full-beam brightness temperature
correction has not been applied (see text).}
\label{subtract_figs_orion.fig}
\end{figure*}

The morphology of extended bright \ha regions can be identified in the
low frequency continuum maps even in the presence of significant
synchrotron emission. By subtracting a template from the radio
continuum maps proportional to the \ha intensity, it is possible to
estimate the amplitude of the free-free emission. A first-order
approach is illustrated for the Orion region in
Fig. \ref{subtract_figs_orion.fig}. The 408 MHz, 2326 MHz, WHAM \ha
and the $D^{T}$ dust maps are first smoothed to 1$^{\circ}$ resolution. Different amplitudes of the free-free emission expected
from the \ha maps are subtracted from the radio maps until the
correlation with the \ha maps disappears from the radio maps; we are
left with the synchrotron features only. Then using the absorption
expected from the $D^{T}$ dust map, an estimate is made of $f_{d}$
for the field. In this case, assuming a nominal electron temperature
of $T_{e}=7000$ K, both the $\lambda$-Orionis nebula $(l,b)
\approx (195^{\circ}, -11^{\circ})$ and
Barnard's Arc give $f_{d}=-0.2 \pm 0.1$, suggesting no dust
absorption. The error bars are estimated by comparing maps with
different amounts of \ha$\!\!$. As shown in Section \ref{scans}, the effect of a lower average
temperature for these regions and/or the full-beam to main-beam
correction, may bring the \ha estimate in agreement
with the radio data ($f_{d}\ge 0)$. 

A similar study was made of the Gum Nebula where $f_{d}$ was found to
be $-0.2 \pm 0.2$, and the Ophiuchus H{\sc ii} region $l \approx
(0^{\circ}) - (20^{\circ})$, $b \approx (+15^{\circ}) - (+30^{\circ})$
where $f_{d} = 0 \pm 0.2$. The increased error in the estimate is
due to the complexity of the emission across each region. In general,
the \ha predictions assuming $T_{e}=7000$ K are correct when a factor
of $0.7-1.0$ is included into the \ha estimates (Fig. \ref{subtract_figs_orion.fig}).

In summary, the raw \ha predictions in these selected regions are slightly
too large to fit the current low-frequency data by up to $\sim
30$ per cent. This may be accounted for by lower than average electron
temperatures (probably the case in some parts of the Orion region)
which will vary the
conversion factor from \ha to radio continuum by up to $\sim 30$ per
cent within a
temperature range of $5000-10000$ K. The full-beam to main-beam
correction may also be a significant correction; up to $50$ per cent, but
$\sim 20-30$ per cent in elongated structures. The resulting $f_{d}$ values for these
regions are consistent with little or no
absorption ($f_{d} \sim 0$). This might be expected since these regions are
at relatively high Galactic latitudes ($|b|>10^{\circ}$) and therefore nearby
objects in which the geometry of the ionization, as seen from the Sun's
position in the Galactic plane, produces ionization on the near side of
the gas/dust clouds.  


\section{The best-estimate free-free template}
\label{free-free_template}

We are now in a position to move from the observed \ha maps of the
northern and southern skies to produce a full-sky dust-corrected \ha map at an
angular resolution of 1\degr. Then, using equation (\ref{tb_ha.eqn}) we can
derive all-sky free-free maps except in areas near the Galactic plane
where the predicted dust absorption becomes uncertain.

\subsection{Full-sky \ha map corrected for dust absorption}
\label{full_ha}

We obtain a full-sky \ha map by using the northern WHAM \ha data at $1^{\circ}$ resolution and the higher
angular resolution Southern SHASSA \ha data. The \ha map produced from
WHAM data was made by
interpolating the irregularly spaced data grid using the Interactive
Data Language (IDL) routines
{\sc triangulate} and {\sc trigrid}, which are based on Delauney triangulation, as
suggested by the WHAM team (L.M. Haffner, private communication). The
median filtered, continuum subtracted composite SHASSA map was used as
given by the SHASSA team (J.E. Gaustad, private communication). Where the maps overlap,
the WHAM map is used because of its higher sensitivity and better
zero-level certainty, except for declinations $<-15^{\circ}$ where
SHASSA data are preferred. The observed maps are regridded onto an
over-sampled Cartesian Galactic coordinate grid and smoothed to
$1^{\circ}$ resolution. We then use the $D^{T}$ map of 100 $\mu$m dust emission to
correct for dust absorption at each point on the \ha map according to
equation (\ref{dust_absorption.eqn}). The fraction of the dust $f_{d}$
lying in front of the \ha is taken to be 0.33 as shown in Section
\ref{latitude_scans} and discussed in Section \ref{obs_tests}.

\begin{figure*}
\vbox to305mm{\vfil \includegraphics[bb=60 120 535 790, angle=90]{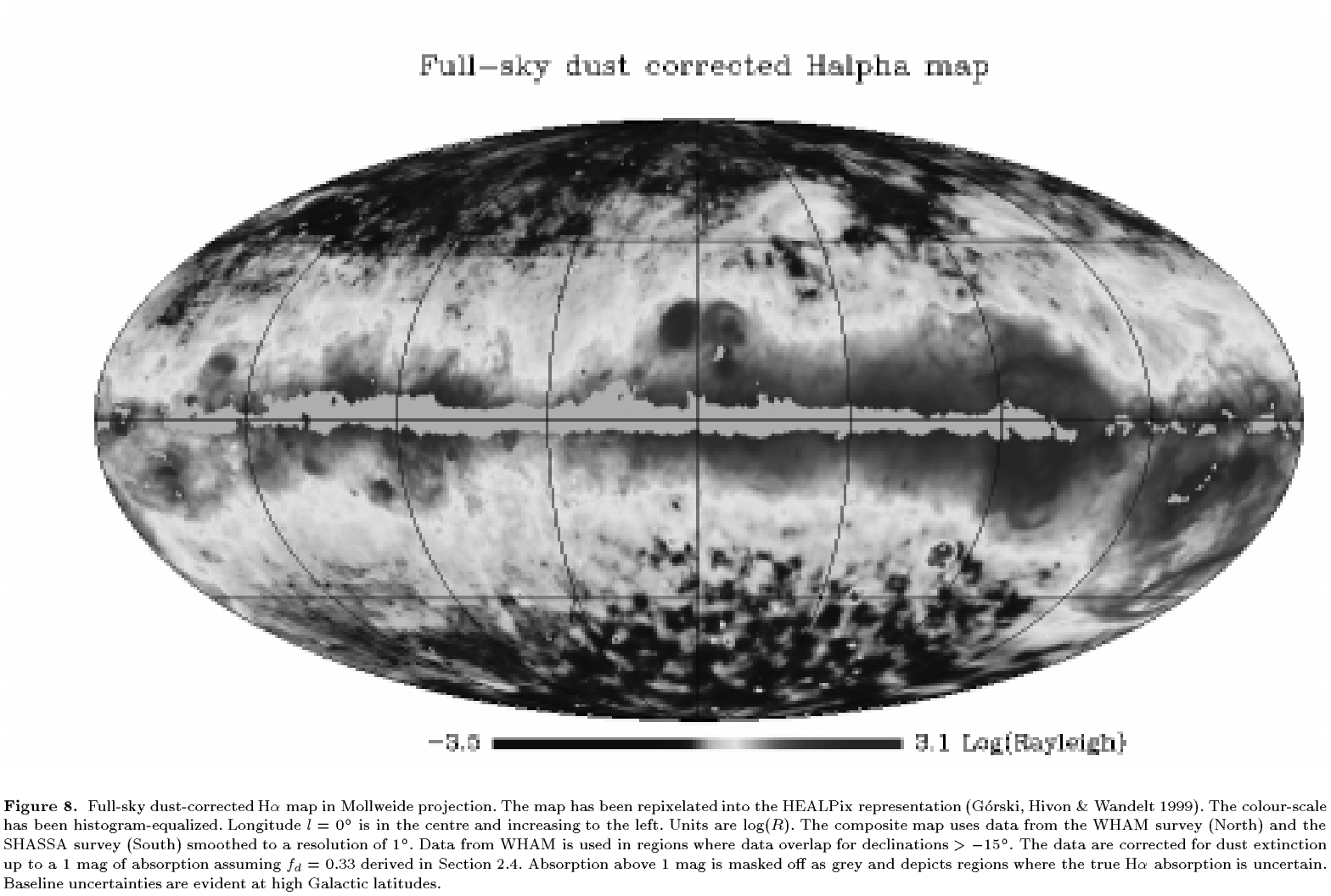}  
\vfil}  
\label{ha_map.fig}
\end{figure*}

Fig. \ref{ha_map.fig} shows a colour representation of the
dust-corrected full-sky \ha map in Mollweide projection. Regions where
the dust absorption correction is greater than 1.0 mag are shown as
grey; \ha data become unreliable in these regions. The Local System (Gould's Belt) is clearly seen beneath the
plane at $l \sim 180^{\circ}$ and above the plane at $l \sim
0^{\circ}$, reaching to $|b| \sim 30^{\circ}$ or more.

It should be remembered that both the WHAM and SHASSA surveys are in
the early stages of analysis and further revisions are
anticipated. Baseline effects are still visible in the data as
presented in Fig. \ref{ha_map.fig}. For example, the WHAM data cannot
accurately subtract the geocoronal emission near the ecliptic pole
($(l,b) \approx (96^{\circ}, 30^{\circ}$)), while the SHASSA data have
artifacts due to the varying geocoronal emission in each $13^{\circ}
\times 13^{\circ}$ field which results in a residual power on these
scales in the region near ($(l,b) \approx (320^{\circ},
-45^{\circ}$)). The typical level of these baseline variations are
$\ltsim 1~R$. 

\subsection{A free-free all-sky template}
\label{full_free}

We can now use Fig. \ref{ha_map.fig} to produce an all-sky free-free
template at any radio frequency using equation (\ref{tb_ha.eqn}). For
illustration we generate the template for 30 GHz, a frequency which is widely
used for CMB studies such as the space projects {\it MAP} and {\it Planck} and
the ground-based arrays CBI, DASI and VSA. In applying equation
(\ref{tb_ha.eqn}) we have taken $T_{e}=7000$ K, the appropriate value
for the region of Galaxy sampled by the \ha maps at $|b| \ga
10$\degr. At 30 GHz, $T_{b}=5.83 ~\mu$K/$R$ (Table \ref{conversions.table}).

The all-sky template for 30 GHz free-free emission is shown in
Fig. \ref{30ghz_template.fig}. The amplitude of the free-free signal
is $< 10 ~\mu$K for $|b|>30^{\circ}$ at most Galactic longitudes, although in the Local System
$T_{b}$ can be as large as 100 $\mu$K. At higher frequencies the
free-free emission falls as $\nu^{-2.15}$ and accordingly at 70 to 100
GHz, where foreground contamination of the CMB is least, the $T_{b}$
values will be 5-10 times less than shown in
Fig. \ref{30ghz_template.fig}. At these frequencies free-free and
vibrational dust emission are the dominant foregrounds. Only the Galactic ridge
($|b| < 5^{\circ}-10^{\circ}$) and isolated regions such as Orion have
$T_{b} > 10-20 ~\mu$K at frequencies $70-100$ GHz. 

We discuss the contribution of free-free emission to the CMB
foregrounds in Section \ref{CMB_compare}.
\setcounter{figure}{8}
\begin{figure*}
\setlength{\unitlength}{1mm}
\begin{picture}(80,92)
\put(0,0){\includegraphics{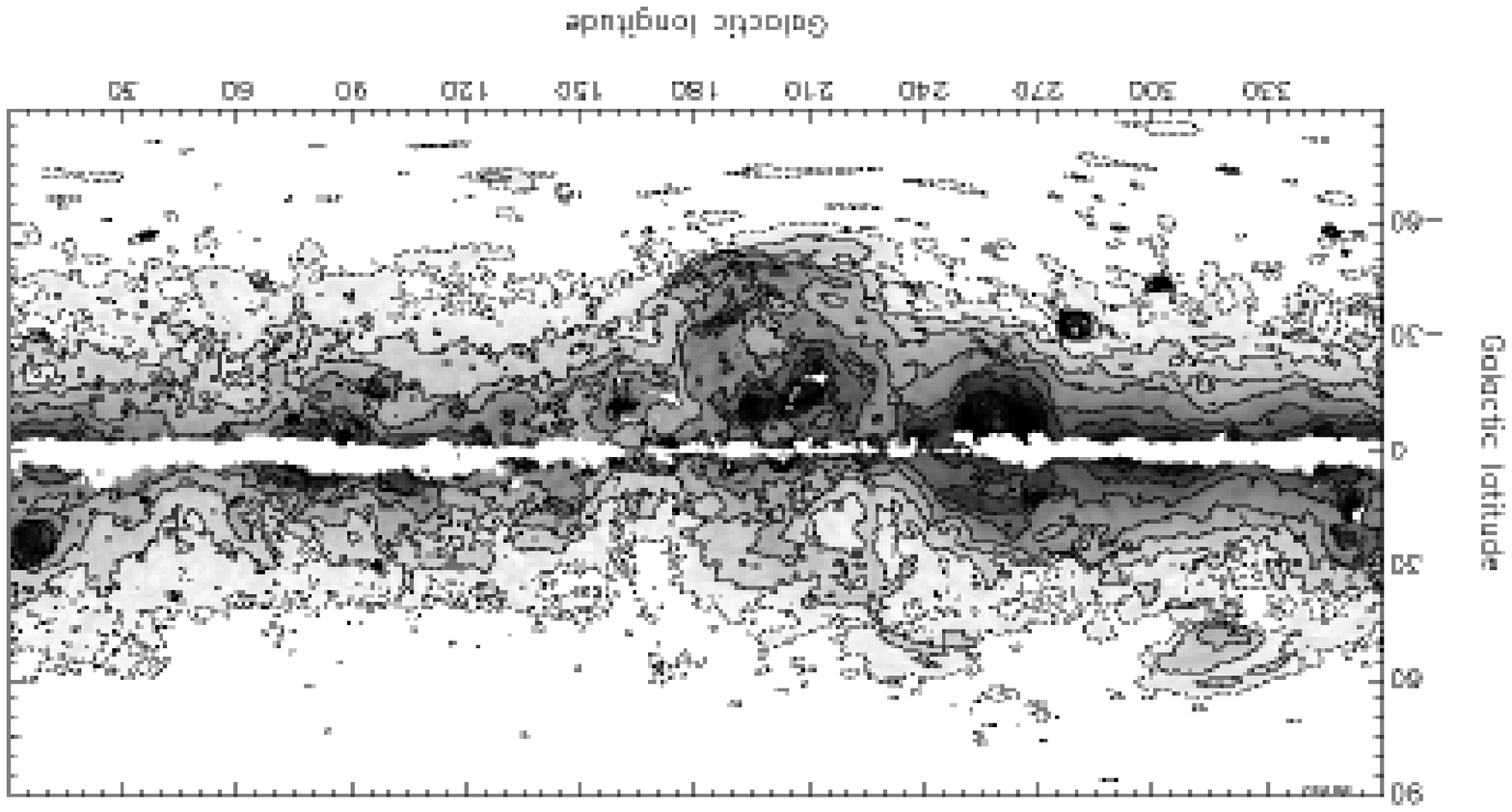}}
\end{picture}
\vskip 3mm
\caption[Free-free template at 30 GHz]{Free-free brightness
temperature template at 30 GHz with $1^{\circ}$ resolution. Grey-scale is
logarithmic from $5-1000~\mu $K. Regions where the template is
unreliable are masked white. Contours are given at 5 (dot-dashed),
10,20,40,100,200 and 500 $\mu$K.} 
\label{30ghz_template.fig}
\end{figure*}


\section{AN IMPROVED LOW-FREQUENCY TEMPLATE}
\label{full_synch}

The 408 MHz all-sky map (Haslam et al. 1982) made at $0^{\circ}\! .85$
resolution is widely used as a template for the Galactic synchrotron
foreground. Our present study allows us to correct this map for
free-free emission at 408 MHz to obtain a clean synchrotron
template. The free-free correction is estimated from the
dust-corrected \ha template of Fig. \ref{ha_map.fig} using
$T_{b}=51.2$ mK/$R$ as given by equation (\ref{tb_ha.eqn}) and Table
\ref{conversions.table} with $T_{e}=7000$ K.

Over much of the sky the free-free features are weaker than the
synchrotron features at 408 MHz. However in regions such as the Local System at
intermediate latitudes and the Gum Nebula at lower latitudes, the
free-free features can be brighter than the synchrotron features. This
is demonstrated in the Orion region in
Fig. \ref{subtract_figs_orion.fig} where significant free-free
emission is seen at both 408 MHz and 2326 MHz. In such
regions the correction is clearly important.

To estimate the level at which free-free emission is present at 408
MHz, the r.m.s. fluctuation of the 408 MHz map smoothed to
$1^{\circ}$ resolution, are compared with that
given by the free-free template map. Table \ref{408mhz_rms.table} gives the results for different
Galactic latitude cuts. The estimate of free-free contamination
suggests that at high Galactic latitude ($|b|>40^{\circ}$), away from
bright \ha features, the free-free component is a few per cent compared with
the synchrotron component. At lower latitudes, the fraction increases
to $\sim 10$ per cent. The all-sky value will be considerably
underestimated since the corrections for dust absorption near the Galactic plane ($|b|<5^{\circ}$) are uncertain. These estimates agree well with previous best
estimates of the ratio of synchrotron and free-free. For example, the
10,15 and 33 GHz Tenerife experiments combined with the 5 GHz
Jodrell Bank interferometer show that the synchrotron and free-free
components at intermediate latitudes are approximately equal at $\sim 10$ GHz (Jones et al. 2001). Assuming spectral indices of $-3.0$ and $-2.1$
for the synchrotron and free-free respectively, the ratio at 408 MHz
extrapolated from 10 GHz predicts $\sim 6$ per cent of 408 MHz emission is in the form
of free-free, in good agreement with the \ha values given in Table
\ref{408mhz_rms.table}.

\begin{table}
\begin{center}
\caption[r.m.s. fluctuations of synchrotron and free-free emission at
408 MHz]{Comparison of the r.m.s. fluctuations at 1$^{\circ}$
resolution of 408 MHz data and the free-free template for different
Galactic latitude cuts. The all-sky value will be a lower limit since
the corrections for dust are uncertain near the Galactic plane ($|b|<5^{\circ}$).}
\begin{tabular}{l|c|c|c}
Galactic &408 MHz &Free-free &Free-free / \\
cut &r.m.s. &r.m.s. &synchrotron \\
 &(K) &(K) &ratio (\%)\\ \hline \hline
All-sky &30.3 &2.9 &$>10$ \\ 
$|b|>10^{\circ}$ &10.3 &0.85 &8 \\
$|b|>20^{\circ}$ &7.2 &0.39 &5 \\
$|b|>30^{\circ}$ &5.9 &0.23 &4 \\
$|b|>40^{\circ}$ &5.1 &0.08 &2 \\ \hline
\label{408mhz_rms.table}
\end{tabular}
\end{center}
\end{table}

Higher frequency maps will necessarily have higher fractions of
free-free emission. The widely used maps at 1420 MHz (Reich \& Reich 1988)
and 2326 MHz (Jonas 1998) are used for estimating the
Galactic contribution at higher frequencies. Using the estimate of $6$
per cent at 408 MHz, the free-free fluctuations at 1420 MHz will contribute
$\approx 15$ per cent  compared to the synchrotron component at
intermediate latitudes, while at 2326 MHz this
increases to $\approx 24$ per cent. By correcting the 408 MHz map for
free-free emission, a better synchrotron map can be made. This is
particularly important for cross-correlation analyses which assume
that the templates are not correlated.

With \ha data alone it is not possible to construct a free-free
correction template at lower Galactic latitudes. Fig. \ref{ha_map.fig} shows
regions along the Galactic plane where the dust absorption is too
large to estimate an accurate dust-free \ha intensity. We plan to use
multi-frequency maps of the Galactic plane to separate the free-free
and synchrotron components at low latitudes. The narrow free-free
latitude distribution along the plane has a peak brightness at
408 MHz which varies from 40 to 80 per cent that of the broader synchrotron
distribution at $l=10^{\circ}$ to $50^{\circ}$ (Large, Mathewson
\& Haslam 1961)
. Accordingly, the correction of the 408 MHz map for free-free
emission is important at low
latitudes.

The other factor in constructing a template at $1^{\circ}$ resolution from
the low frequency radio maps is the effect of the main-beam to
full-beam ratio ($\sim 1.5$) on angular scales of a few degrees; we
note that the brightness temperature scale of the published maps is on
the full-beam scale ($\sim 5^{\circ}-10^{\circ}$). The result is that
temperatures on the main-beam scale ($\sim 1^{\circ}$) should be
multiplied by a factor $\sim 1.5$ to bring them to the correct brightness
temperature scale. This procedure has been considered in Sections \ref{scans}
and \ref{subtract} and clearly needs to be taken into account for
accurate free-free and synchrotron predictions. This scale-dependence to true
brightness temperature has been noted before (e.g. Jonas et al. 1998)
and is currently under review when applying it to existing full-sky maps.

\section{DISCUSSION}
\label{discussion}

\subsection{Precision of the free-free template}

The \ha intensities given by WHAM and SHASSA have quoted calibration
accuracies of 10 and 9 per cent respectively. The corresponding zero
level uncertainties are approximately 0.2 and 0.5 $R$. The
accuracy is likely to be improved as these surveys are refined. The
largest uncertainty in the \ha intensity is in the dust
absorption correction as discussed in Section \ref{dust_absorption}. At intermediate and higher Galactic latitudes where $f_{d} \times
A_{{\rm H}\alpha} \le 0.2$ mag, the error in the corrected $I_{{\rm
H}\alpha}$ is $\le 5$ per cent; this error increases to $\sim 30$ per
cent where $f_{d} \times A_{{\rm H}\alpha}$ is 1 mag nearer the
Galactic plane.

On converting the dust-corrected $I_{{\rm H}\alpha}$ to $T_{b}$, the
free-free brightness temperature, a further uncertainty is introduced
by the spread in $T_{e}$, the electron temperature, in the local
region of the Galaxy covered by the \ha maps. For $T_{e}=7000 \pm
1000$ K, the $T_{b}$ uncertainty is $\pm 10$ per cent as derived from
equation (\ref{tb_ha.eqn}). As far as the emission theory used in
Section \ref{halpha_conversion} is concerned, we believe that the Gaunt
factor and the analytical relations given are accurate to 1-2 per
cent. It is assumed throughout that the \ha emission is given by the
case B emission formulae; the observed radio $T_{b}$ values are
consistent with this assumption.
\begin{figure*}
\setlength{\unitlength}{1mm}
\begin{picture}(80,60)
\put(0,0){\includegraphics{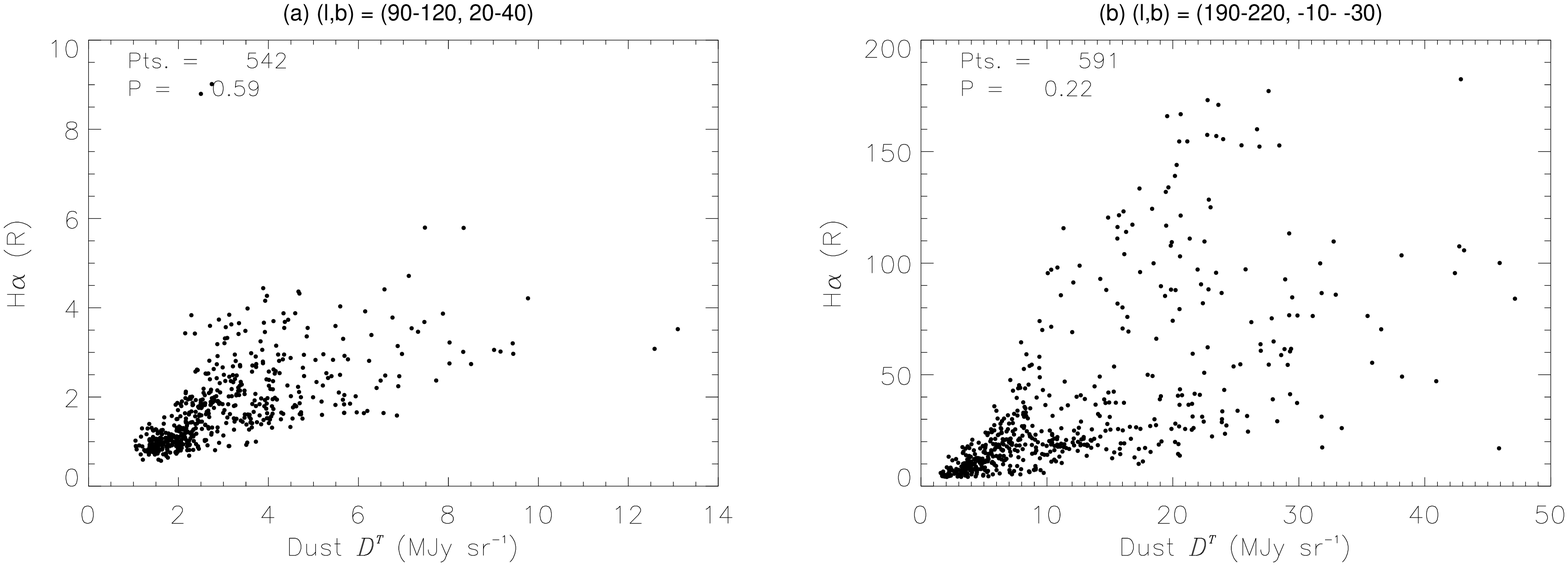}}
\end{picture}
\vskip 3mm
\caption[Correlation plots between dust and \ha]{Correlation plots
between \ha and dust at a resolution of $1^{\circ}$ for two regions (a)
$l=(90^{\circ}-120^{\circ}), b=(20^{\circ}-40^{\circ})$ and (b) $l=(180^{\circ}-210^{\circ}),
b=(-20^{\circ}$ to $-40^{\circ})$. There is an overall positive
correlation with considerable scatter. The Orion region (b) is
brighter and shows a much larger scatter. The Pearson's correlation
coefficients are 0.59 and 0.22 respectively.}
\label{TT_plots.fig}
\end{figure*}

\subsection{The contribution to other templates}

A radio-independent method of determining a free-free template has a
significant benefit in establishing the other Galactic foreground
templates for CMB studies. We have already shown in Section
\ref{full_synch} that an improved synchrotron template can be derived
from the 408 MHz map by correcting for the known free-free
emission. This method breaks down for the small area along the
Galactic plane where the dust absorption of the \ha becomes too high;
other methods of determining the free-free emission are available
here.

\begin{figure*}
\setlength{\unitlength}{1mm}
\begin{picture}(80,80)
\put(0,0){\includegraphics{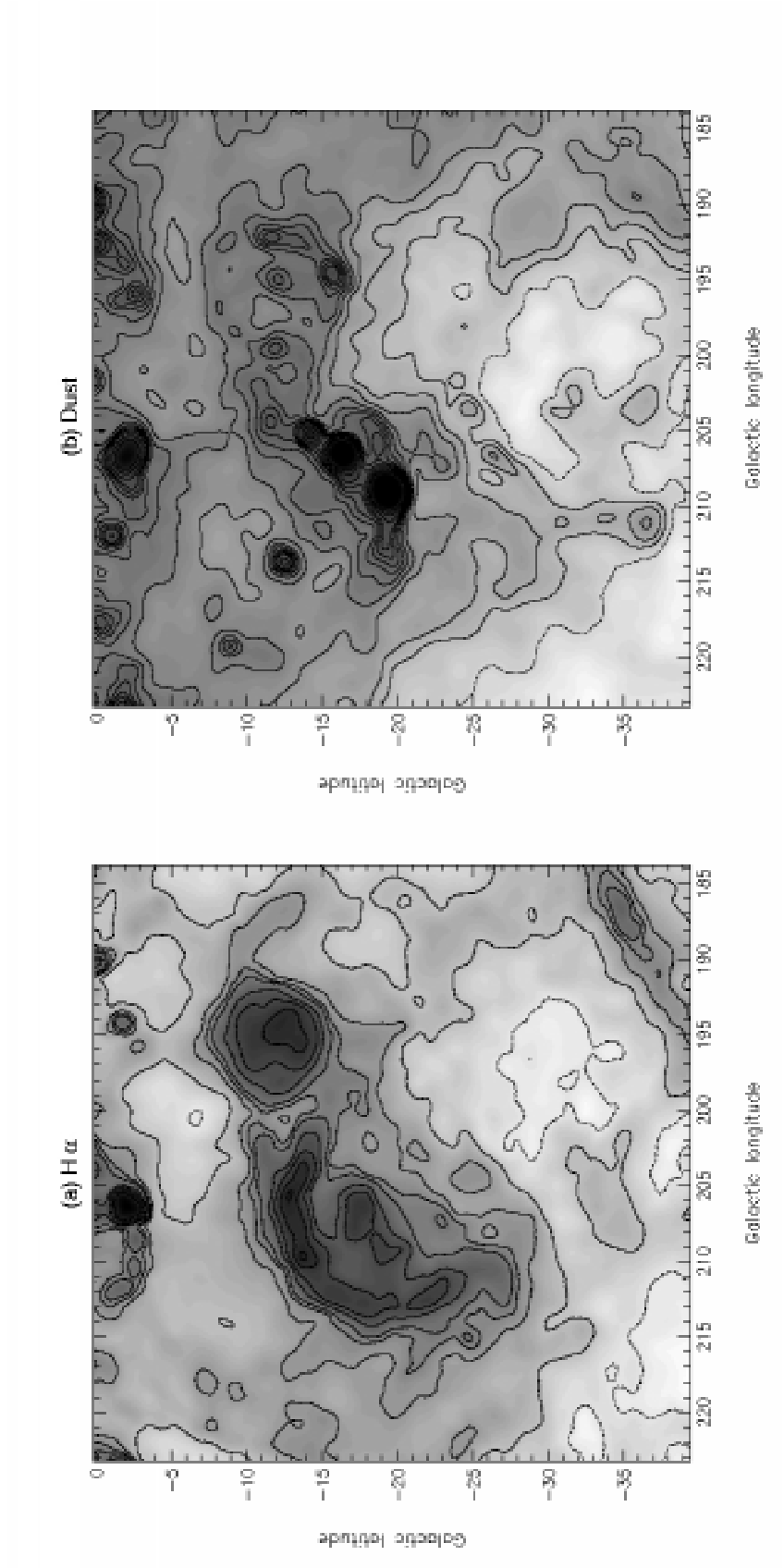}}
\end{picture}
\vskip 3mm
\caption[Correlations between dust and \ha]{Correlations between \ha
and dust in the Orion region. (a) \ha data from WHAM shown as a
logarithmic grey-scale from 3 to 380 $R$ and contoured at 5
(dot-dashed),10,20,30,40,70,100,150 and 200 $R$. (b) 100 $\mu$m data from
SFD98 depicting the dust in a logarithmic grey-scale from 0.1 to 200 MJy
sr$^{-1}$ and contoured at 3
(dot-dashed),5,7,10,15,20,25,30,35,40,50,70 and 100 MJy
sr$^{-1}$. There is an overall correlation between \ha and dust but it
is not one-to-one.}
\label{correlations.fig}
\end{figure*}

Spinning dust radiating in the region 10-40 GHz is proposed as a
Galactic foreground (Draine \& Lazarian 1998); observations appear
to confirm this scenario (Kogut et al. 1996; Leitch et al. 1997; de Oliveira-Costa et
al. 1997, 1998, 1999, 2000, 2002; Finkbeiner et al. 2002). A strong confirmation of the
presence of spinning dust radio emission correlated with the SFD98
dust template requires a clean separation of the free-free component
beforehand. This is important because the dust and \ha are generally
thought of as being correlated. An example of a region of the Galaxy
where the correlation is relatively strong is
$l=(90^{\circ}-120^{\circ}), b=(20^{\circ}-40^{\circ})$. This is
illustrated in Fig. \ref{TT_plots.fig}(a) which shows the correlation
between \ha and ($D^{T}$) from SFD98 at $1^{\circ}$ resolution. The relation between $I_{{\rm H}\alpha}$ and
$D^{T}$ for the region in Orion $l=(180^{\circ}-210^{\circ}),
b=(-20^{\circ}$ to $-40^{\circ})$ is given in Fig. \ref{TT_plots.fig}
(b) where it is seen
that the correlation is less strong. The distribution of \ha and dust
for the Orion region is shown in Fig. \ref{correlations.fig}. The situation is clearly complex on the scale of $1^{\circ}$ as
would be expected for a star-forming region and underscores the
necessity of removing the free-free emission before investigating
spinning dust.

\subsection{Free-free contribution to the CMB foregrounds}
\label{CMB_compare}

We will now estimate the free-free contribution to
the CMB foregrounds at $\sim 30$ GHz. If
not subtracted from the data, the
foreground signal will add in quadrature to the CMB signal in the
power spectrum since they will be uncorrelated on the sky. The free-free template allows
statistical estimates to be made of the free-free component on angular
scales $\gtsim 1^{\circ}$ compared to the primordial CMB
fluctuations. At angular scales of $1^{\circ}$ ($\ell \approx 200$),
the CMB has r.m.s. fluctuations of $\approx 75~\mu$K, falling to
$\approx 30~\mu$K at smaller and larger angular scales. Over a large
portion of the higher latitude sky ($|b|>40^{\circ}$), the average contribution of
free-free emission is $\approx 9~\mu$K at 30 GHz on scales of $1^{\circ}$. This
amounts to a $1$ per cent increase in the power spectrum at
$\ell=200$. However, individual regions may be contaminated at a much higher
level depending on their position. At higher frequencies ($\sim
70-100$ GHz), the free-free contribution will be  $\ltsim 1~\mu$K and
will be negligible at high latitudes.

We can also compare the estimates for specific regions of the sky where CMB
observations have been made. The VSA has observed three regions of the
sky measuring angular scales of $\approx 0^{\circ}\!.3-2^{\circ}\!.0$
($\ell \approx 150-900$) covering a total area of 101 square degrees at a frequency of 34
GHz (Taylor et al. 2002). The contamination from free-free emission
using the free-free template is estimated to give an
r.m.s. temperature of $0.6-0.8~\mu$K at the $1^{\circ}$ scale and hence the free-free contamination is
negligible in these selected regions. 

The North Celestial Pole (NCP) region has been studied by the
Saskatoon group (Wollack et al. 1997). The Saskatoon data show an
anomalous component which cannot be explained on conventional
grounds. Using the free-free template, the r.m.s. variations are
$\approx 3.4~\mu$K in a $15^{\circ}$-diameter circle centred on the
NCP on smoothing the \ha map to a resolution of $1^{\circ}\!.45$ and
assuming a conversion factor of $\approx 7~\mu$K$/R$. Simonetti et al. (1996) use their \ha data to estimate an upper limit of
$4.6~\mu$K at 27.5 GHz in this region. These values of $\sim 4\mu$K can
be compared with the $\approx 40~\mu$K r.m.s. variations in the
Saskatoon data.  It therefore seems clear,
that the anomalous emission is not in the form of free-free emission
from gas at $T_{e} \sim 10000$ K. The current view is
that it may be due to spinning dust emission as proposed by Draine \&
Lazarian (1998).


\section{CONCLUSIONS}
\label{conclusions}

The recent publication of \ha surveys covering the majority of the sky has
provided a break-through in obtaining a free-free Galactic foreground
template of importance for CMB and ISM studies. Having derived the
free-free template it is then possible to determine a better
synchrotron template and to make real progress in constructing a
spinning dust template. The free-free template has many other
applications which include a re-analysis of the {\it COBE-DMR} data (Banday et al., in
preparation) and a free-free power spectrum analysis (Dickinson et al.,
in preparation).

We can look forward to an improved all-sky free-free template at
higher resolution than the $1^{\circ}$ used in the present work. A
combination of the high resolution (arcmin scale) narrow-band filter
surveys such as SHASSA and VTSS combined with the high sensitivity Fabry-Perot
survey at a $1^{\circ}$ scale would be ideal. At lower Galactic latitudes
where the dust absorption of the \ha emission is significant,
multi-frequency continuum surveys  should be able to separate the
strong synchrotron and free-free components. Recombination line
surveys such as those being undertaken with HIPASS in the South
(Barnes et al. 2001) and HIJASS in the North (Kilborn 2002) will
provide a confirmation of the low latitude free-free template and
distinguish it from spinning dust which will have a very similar
latitude distribution. The recombination line data will also lead to a kinematic
2-dimensional picture of the ionized gas distribution in the Galaxy
using kinematic distances since they contain velocity information.

\subsection*{Acknowledgments}

CD acknowledges a PPARC research grant. WHAM is funded by the National
      Science Foundation (NSF). The Southern H-Alpha Sky Survey Atlas
      (SHASSA) is supported by the National Science Foundation. We
      thank the WHAM/SHASSA teams for making their surveys available and answering
      our queries. We also thank David Valls-Gabaud for verifying the
      calculation of the Gaunt factor. 

\bibliographystyle{mnras}

\label{lastpage}
\end{document}